\documentclass[article,preprintnumbers,nofootinbib,showpacs,notitlepage,unsortedaddress,superscriptaddress,longbibliography]{revtex4}
\usepackage{amsmath}
\usepackage{verbatim}
\usepackage{psfrag}
\usepackage{bm}
\usepackage{bbm}
\usepackage{graphicx}
\usepackage{latexsym}
\usepackage[english]{babel}
\usepackage{amsthm}
\usepackage{color}
\usepackage{amssymb}

\begin{document}



\vskip .5cm

\title{\Large The Gaussian entropy of fermionic systems }

\preprint{ITP-UU-12/13, SPIN-12/11, HD-THEP-12-1}


\author{Tomislav Prokopec}
\email[]{T.Prokopec@uu.nl} \affiliation{Institute for Theoretical
Physics (ITP) \& Spinoza Institute, Utrecht University, Postbus
80195, 3508 TD Utrecht, The Netherlands}

\author{Michael G. Schmidt}
\email[]{M.G.Schmidt@thphys.uni-heidelberg.de} \affiliation{
Institut f\"ur Theoretische Physik, Heidelberg University,
Philosophenweg 16, D-69120 Heidelberg, Germany}

\author{Jan Weenink}
\email[]{J.G.Weenink@uu.nl} \affiliation{Institute for Theoretical
Physics (ITP) \& Spinoza Institute, Utrecht University, Postbus
80195, 3508 TD Utrecht, The Netherlands}

\begin{abstract}

We consider the entropy and decoherence in fermionic quantum systems.
By making a Gaussian Ansatz for the density operator of a collection
of fermions we study statistical 2-point correlators and
express the entropy of a system fermion in terms of these correlators.
In a simple case when a set of $N$ thermalised environmental fermionic
oscillators interacts bi-linearly with the system fermion we can study
its time dependent entropy, which also represents a quantitative measure
for decoherence and classicalization.
We then consider a relativistic fermionic quantum field
theory and take a mass mixing term as a simple model for the Yukawa
interaction.
It turns out that even in this Gaussian approximation, the fermionic system
decoheres quite effectively, such that in a large coupling and high temperature
regime the system field approaches the temperature of the environmental fields.

\end{abstract}

\maketitle

\section{Introduction}
\label{Introduction}

 The density operator contains complete information about quantum statistical
systems, and hence it can be used to study various properties
of such systems, such as correlators, particle numbers,
entropy and decoherence. However, the evolution of realistic physical
systems is governed by interacting field theories, and only rarely
the density operator is known beyond a perturbative approximation,
which can be, nevertheless, very useful for weakly coupled regimes.

 Even the Gaussian part of the density operator contains
important information about the entropy and decoherence of the system,
which can be neatly encoded in the statistical two point function,
as was firstly pointed out independently by two groups of
authors~\cite{Giraud:2009tn,Koksma:2009wa}. This correlator approach to
entropy, decoherence and classicalization has been extensively used
in the context of weakly interacting bosonic systems
\cite{Calzetta:1986cq,Calzetta:2003dk,Campo:2008ju,Campo:2008ij,
Giraud:2009tn,Koksma:2009wa,Koksma:2010dt,Koksma:2010zi,Koksma:2011fx,Koksma:2011dy}.
However very little is known about the entropy and decoherence
in fermionic systems, and the corresponding literature
is scarce~\cite{Floreanini:1987gr,Kiefer:1993fw,Benatti:2000wu}.
In this paper we present a first study of entropy and decoherence
in relativistic fermionic field theories. For simplicity, we consider
here only simple bilinear interactions, which are in field theory
known as mass mixing. Since our Hamiltonian is quadratic in the fields,
an initial Gaussian density operator will remain Gaussian
as the system evolves, and a complete information about the density operator can
be given in terms of equal-time $2$-point correlators, which is the strategy
we use in this work. For pedagogical reasons, we begin by considering
coupled fermionic quantum oscillators, and only then move on to field theory.

If the density operator of a system $\hat\rho$ is known,
the (information) entropy $S_{\rm vN}$ can be calculated by the von Neumann formula,
 \begin{equation}
   S_{\rm vN} = - \langle \ln(\hat\rho)\rangle
               = - {\rm Tr}[\hat \rho\ln(\hat \rho)]
\,.
\label{entropy:vN}
\end{equation}
Now, by making use of the Heisenberg evolution equation for the density
operator, one can easily show that the von Neumann entropy is conserved for
closed systems. In practice however no observer $O$ will have access to a
complete information of any nontrivial system $S$ (with many interacting
degrees of freedom), making the system open. Such systems will interact with
an environment $E$ which is, by definition, inaccessible to $O$.
The loss of information associated with this inaccessibility generically
leads to decoherence~\cite{Zeh:1970,Zurek:1981xq,Joos:1984uk,joos2003decoherence,Zurek:2003zz},
a rather qualitative concept that
describes how a system evolves into a state which most closely resembles
a classical state. However, at the same time
the (reduced) von Neumann entropy~\eqref{entropy:vN}
of the system alone~\footnote{The von Neumann  entropy of the reduced system
density operator is in literature also known as the entanglement
entropy~\cite{Eisert:2008ur,Solodukhin:2011gn},
as the information about the entanglement between the system and environment
is lost in the reduced density matrix.}
is no longer conserved due to this loss of information. Entropy generation thus
provides a quantitative measure of decoherence and classicalization.

 This statement requires clarification,
because decoherence is an observer-dependent concept, whereas
entropy \eqref{entropy:vN} can be defined without introducing an observer,
but of course it does depend on the system -- environment split.
To make our work as general as possible, we kept the observer implicit
throughout. Yet, a natural observer $O$ is the one which
entangles with those states that diagonalise the density operator.
In this work we identify those states as the Fock states for
which statistical (average) particle number is defined. Decoherence
is now induced by tracing over the inaccessible environmental
degrees of freedom, and is perfect from the viewpoint of the observer
entangled with the Fock states
that does not see the entanglement between system and environment.
As these Fock states get more occupied,
and the system's entropy increases, the system gets more classical.
This observer plays a special role in the class of all observers
sensitive to Gaussian properties of the density operator,
and that is that this observer sees the most quantum properties of the system.
All other observers, such as the momentum or position operator,
will perceive the system as more classical. We can make this more concrete
by considering the example of a highly squeezed pure state
(that describes {\it e.g.} linearized cosmological perturbations at the end of inflation).
That state will be perceived as classical by the observer that measures
the position spread of the state (in the sense that
$\Delta x$ will be much greater than in
the pure vacuum state), but it will be perceived as a pure quantum state by the
observer that entangles with the Fock states that diagonalize the system's
density operator.

 But, what can be the inaccessible information that can be justifiably called
an environment? The most common example is a thermal bath of particles
interacting weakly with the system that is observed.
In this case, the system-environment correlations -- also known as
entanglement -- are not observable.
These correlations can be in $2$-point $SE$ correlations (such as considered
in this work) or in higher order $n$-point functions.
Higher order correlators are always suppressed by
some power of the coupling constant, and hence they are typically small
in a weakly coupled regime. For example, in
Refs.~\cite{Koksma:2009wa,Koksma:2011fx} the leading $SE$ correlator
that is neglected corresponds to a $3$-point $SE$ correlator.

 In this work we take the interaction to be of a Yukawa type,
schematically
\begin{equation}
 {\cal L}_{\rm int}=-y_{ij}\hat{\bar\psi}_i\hat \phi\hat\psi_j
\,,
\label{Yukawa}
\end{equation}
where $\hat\phi$ and $\hat\psi_j$ denote a scalar and fermionic quantum field,
respectively. The simplest approximation in which one can treat
this interaction is to neglect quantum fluctuations of the scalar field,
{\it i.e.} to replace the scalar field by its expectation value,
\begin{equation*}
 \hat\phi\rightarrow \phi\equiv {\rm Tr}[\hat \rho\hat\phi]
\,.
\end{equation*}
Within this Gaussian approximation the Yukawa interaction reduces
to a mass mixing term,
\begin{equation}
 {\cal L}_{\rm int}\rightarrow - \hat{\bar\psi}_iM_{ij}\hat\psi_j
\,,\qquad
M_{ij}=y_{ij}\phi
\,,
\label{Yukawa:gaussian}
\end{equation}
the Hamiltonian becomes quadratic in the fields,
and the problem becomes exactly soluble.
We shall use numerical techniques to obtain an exact solution to this
simplified problem, and we shall express
the Gaussian density operator in terms of equal-time 2-point correlators.
The information we consider inaccessible to $O$ is in
the $SE$ and $EE$ $2$-point correlators, and hence entropy gets generated.
The so-called {\it Gaussian} von Neumann entropy for the system field,
which is derived from the Gaussian density matrix alone, yields
a good quantitative measure of decoherence for nearly Gaussian systems.
However, for highly non-Gaussian
systems, one has to modify the entropy definition to incorporate the relevant
non-Gaussian features of the state~\cite{Koksma:2010zi}.
In the simplified problem \eqref{Yukawa:gaussian} the Gaussian von Neumann entropy
can be analytically calculated
in terms of 2-point (statistical) correlators of the system degrees of freedom.
This has also been done for various bosonic systems in
Refs.~\cite{Koksma:2009wa,Koksma:2010dt,Koksma:2010zi}. In this work we set out to derive
the Gaussian (von Neumann) entropy for fermionic systems in terms
of correlators of the system.

The Gaussian entropy is generally not conserved in the presence
of interactions, which could be either environmental interactions,
or self-interactions. Several case studies for bosonic
systems \cite{Koksma:2010dt,Koksma:2009wa,Koksma:2010zi,Koksma:2011dy,Giraud:2009tn}
have indeed shown that the Gaussian entropy increases
for interacting systems, thereby quantitatively describing decoherence
and classicalization. As far as we know, a quantitative description
of decoherence for
fermionic systems is still lacking. A better understanding of
decoherence in fermionic systems can be applicable in many situations of
physical interest. For instance,
one species of fermions could mix with others through mass-type terms,
such as quarks through the CKM matrix~\cite{Cabibbo:1963yz,Kobayashi:1973fv}
or neutrinos through the PMNS~\cite{Pontecorvo:1967fh,Maki:1962mu} matrix.
Other examples include Yukawa interactions, or condensed matter systems
with interacting fermions. Here a framework is provided for
calculating the growth of entropy for a system of fermions
interacting {\it via} a fermionic mass matrix~\eqref{Yukawa:gaussian}.
As explained above, this model represents the simplest (Gaussian)
approximation to the more realistic Yukawa interaction~\eqref{Yukawa}.

The outline of this work is as follows: in
section~\ref{Fermionic quantum mechanics as a 1+0 dimensional field theory}
the simplest example of a one-dimensional fermionic harmonic oscillator
is discussed. A general {\it Ansatz} is made for the density operator,
after which the particle number and entropy
are derived in terms of the statistical correlators of the system.
In section~\ref{Coherent states} we make a connection with some existing
literature by working with the density operator in the coherent state basis.
Next, in section~\ref{sec:Fermionic dynamics} the simplest possible
interactions are added to the fermionic system: $N$ environmental fermionic
oscillators coupled bilinearly to the system oscillator. Though not
completely realistic, this example provides an insight into how
a loss of information leads to an increase in the entropy of a fermionic system,
and some specific examples are shown.
In section~\ref{sec:Entropy generation in a fermionic quantum field theory}
we switch our attention to the more realistic fermionic quantum field theory.
After discussing diagonalisation of the Dirac Hamiltonian in~\ref{sec: hamiltonian diagonalisation},
an {\it Ansatz} is made for the density operator in terms of
mixing particle and antiparticle states in section~\ref{sec: qft density operator entropy}.
The Gaussian entropy is derived in terms of the statistical correlators.
In section~\ref{sec: Ndegrees of freedom}
the fermionic entropy is generalised in the presence
of $N$ fermionic degrees of freedom.
Finally, in section~\ref{sec: fermion mass mixing} the first realistic
example of entropy generation in fermionic quantum field theory is discussed,
which is the simple case of one fermionic species mixing with other species
through mass terms.

\section{Entropy generation in fermionic quantum mechanics}
\label{Fermionic quantum mechanics as a 1+0 dimensional field theory}

 The most general {\it Ansatz} for the density operator
of a free fermionic quantum mechanical system (fermionic harmonic oscillator)
with the Lagrangian,
\begin{equation}
 L_0 
= \hat\psi^\dagger (\imath \partial_t
           -\omega(t))\hat\psi
\,,
\label{HO:lagrangian}
\end{equation}
can be written as
\footnote{
The anticommutation relation~(\ref{canonical anticommutation}) implies
that the other possible Gaussian term $\exp(-b\hat\psi\hat\psi^\dagger)$
can be expressed in terms of $\exp(-a\hat\psi^\dagger\hat\psi)$
plus an appropriate change in the normalisation constant,
and hence does not constitute a new term.
In presence of interactions, the lagrangian $L_\psi$ can be written as
$L_\psi =  L_0 + L_{\rm int}$, where
$L_{\rm int} =  - \hat j_\psi^\dagger\hat\psi- \hat\psi^\dagger\hat j_\psi$.
In this case the density operator can still
be written as in~\eqref{HO:rho:Ansatz}, where now
$\hat\psi$ and $\hat\psi^\dagger$ denote the suitably shifted fields,
as shown in appendix~\ref{Appendix B: Fermionic shift and diagonalisation}.
},
\begin{equation}
 \hat \rho(t) = \frac{1}{Z}\exp(-a\hat\psi^\dagger\hat\psi)
 \,,
\label{HO:rho:Ansatz}
\end{equation}
where $a(t)$ is a (complex valued) function of time, and
$1/Z$ is the normalisation constant determined by the usual trace condition,
\begin{equation}
 {\rm Tr}[\hat\rho(t)]=1
\,,
\label{normalisation}
\end{equation}
and $\hat\psi$ is the (Grassmannian) fermionic operator
(here expressed in the Schr\"odinger picture) satisfying
the usual canonical anticommutation relation,
\begin{equation}
\{\hat\psi,\hat\psi^\dagger\}=1
\,.
\label{canonical anticommutation}
\end{equation}
Now making use of the Grassmannian nature of the operators $\psi^\dagger$
and  $\psi$ with $(\hat\psi)^2=0$,
$(\hat\psi^\dagger)^2=0$, and of~\eqref{canonical anticommutation}, we can
expand~\eqref{HO:rho:Ansatz} as,
\begin{align}
\hat \rho(t) = \frac{1}{Z}\left(1+\left[{\rm e}^{-a}-1\right]\hat N\right)
\,,
\label{HO:rho:expand:2}
\end{align}
where we introduced the fermionic number operator
$\hat N =\hat\psi^\dagger\hat\psi$ with $\hat N^n = \hat N$ ($n=1,2,..$).
The Hilbert space of this theory is two dimensional, and can
be conveniently represented in terms of the Fock space basis vectors
$\{|0\rangle,|1\rangle\}$, defined by,
\begin{equation}
\hat N|n\rangle = n|n\rangle
\,.
\label{Fock basis}
\end{equation}
The trace of the density operator~\eqref{normalisation} is easily evaluated
in this basis,
\begin{equation}
 {\rm Tr}[\hat\rho(t)]=\frac{1}{Z}\sum_{n=0,1}
         \langle n|\left(1+\Big[{\rm e}^{-a}-1\Big]\hat N\right)|n\rangle
        \,,
\label{normalisation:2}
\end{equation}
such that the general Gaussian fermionic density operator is properly normalised
according to \eqref{normalisation} by $Z=1+\exp(-a)$.
It is also convenient to express the density operator as
\begin{equation}
 \hat \rho(t) = (1-\bar{n}(t))+(2\bar{n}(t)-1)\hat N
\,,
\label{HO:rho:solution:2}
\end{equation}
where the average particle number $\bar n$ is defined as,
\begin{equation}
\langle \hat N\rangle = {\rm Tr}[\hat{\rho}(t) \hat N]
                      = \frac{1}{{\rm e}^{a}+1}\equiv \bar{n}(t)
\label{average particle number}
\,.
\end{equation}
The (von Neumann) entropy is then simply,
\begin{align*}
S &= -{\rm Tr}[\hat \rho\ln(\hat \rho)]
\\
  &= -\sum_{n=0,1}\langle n|
        \left\{
           \big[(1-\bar n)+(2\bar n-1)\hat N\big]
               \ln\big[(1-\bar n)+(2\bar n-1)\hat N\big]
         \right\}
       |n\rangle
\,.
\end{align*}
This evaluates to
\begin{equation}
 S = -(1-\bar n)\ln(1-\bar n)-\bar n\ln(\bar n)
\,,
\label{HO:entropy}
\end{equation}
which is the standard expression for the entropy of
$\bar n$ free (non-interacting) fermions, where
$\bar n$ is the average number of fermions in the system defined
in~\eqref{average particle number}. For an analogous discussion of
a bosonic oscillator we refer
to appendix~\ref{Appendix A: Bosonic density operator and entropy}.

 Let us now make a connection with the familiar expressions for
a thermal fermionic density matrix~\cite{kubo1990statistical}.
According to the Fermi-Dirac distribution, the average
occupancy of a state with energy $E$ is given by,
\begin{equation}
\bar n_{\rm FD} = \frac{1}{{\rm e}^{\beta E}+1}
\,,
\label{Fermi Dirac}
\end{equation}
where $\beta=1/(k_B T)$ is the inverse temperature.
Of course,
\begin{equation*}
\bar n_{\rm FD}\equiv\langle \hat N\rangle
           = {\rm Tr}[\hat \rho_{\rm th}\hat N]
\,,
\end{equation*}
where $\hat \rho_{\rm th}$ denotes a thermal density operator.
By comparing with the general expression for
$\hat \rho$~\eqref{HO:rho:solution:2}
and with~\eqref{average particle number}, it is now easily seen
that the thermal density matrix is obtained upon identification,
$a\rightarrow \beta E$, such that,
\begin{equation}
 \hat \rho_{\rm th} = \frac{1}{{\rm e}^{\beta E}+1}
         \left({\rm e}^{\beta E}+\Big[1-{\rm e}^{\beta E}\Big]\hat N\right)
\,,
\label{thermal density operator}
\end{equation}
or, equivalently,
\begin{equation}
 \hat \rho_{\rm th} = (1-\bar n_{\rm th})+(2\bar n_{\rm th}-1)\hat N
\,,
\label{thermal density operator:2}
\end{equation}
which is, as expected,
of the same form as the general entropy~\eqref{HO:entropy}.
The thermal density operator~\eqref{thermal density operator:2}
 implies the following well known expression
for the entropy of a thermal Fermi gas,
\begin{equation}
 S_{\rm th} = -(1-\bar n_{\rm th})\ln(1-\bar n_{\rm th})
        -\bar n_{\rm th}\ln(\bar n_{\rm th})
\,.
\label{thermal entropy}
\end{equation}

\medskip

Let us now consider a bit more closely Eq.~(\ref{HO:entropy}).
In the spirit of the Schwinger-Keldysh out-of-equilibrium formalism,
it is convenient to introduce the following 2-point functions,
\begin{align}
 \imath S^{++}(t;t^\prime)
   &= \;\;\langle T[\hat\psi(t)\hat\psi^\dagger(t^\prime)]\rangle
\nonumber\\
 \imath S^{+-}(t;t^\prime)
   &= -\langle \hat\psi^\dagger(t^\prime)\hat\psi(t)\rangle
\nonumber\\
 \imath S^{-+}(t;t^\prime)
   &= \;\;\langle \hat\psi(t)\hat\psi^\dagger(t^\prime)\rangle
\nonumber\\
 \imath S^{--}(t;t^\prime)
   &= \;\;\langle \bar T[\hat\psi(t)\hat\psi^\dagger(t^\prime)]\rangle
\,,
\label{2point functions}
\end{align}
where here $\hat \psi(t)$ and $\hat\psi^\dagger(t)$ ade the Heisenberg
picture operators, and
$T$ and $\bar T$ denote time ordering and anti-time ordering operations,
defined as,
\begin{align}
 \imath S^{++}(t;t^\prime)
   &= \theta(t-t^\prime) \imath S^{-+}(t;t^\prime)
       + \theta(t^\prime-t) \imath S^{+-}(t;t^\prime)
\nonumber\\
 \imath S^{--}(t;t^\prime)
   &= \theta(t-t^\prime) \imath S^{+-}(t;t^\prime)
       + \theta(t^\prime-t) \imath S^{-+}(t;t^\prime)
\,,
\label{2point functions:Tordered}
\end{align}
such that,
\begin{equation*}
  \imath S^{++} + \imath S^{--} = \imath S^{+-} + \imath S^{-+}
\,.
\end{equation*}
 The retarded and advanced Green functions are then,
\begin{equation*}
 \imath S^{\rm r} = \imath S^{++} -\imath S^{+-}
                  = -(\imath S^{--} -\imath S^{-+})
\,,\qquad
 \imath S^{\rm a} = \imath S^{++} -\imath S^{-+}
                  = -(\imath S^{--} -\imath S^{+-})
\,.
\end{equation*}
Notice that $\imath S^{\rm r}$ and $\imath S^{\rm a}$ can be also written
as
\begin{equation}
 \imath S^{\rm r}(t;t^\prime) = \theta(t-t^\prime)
               \langle\{\hat\psi(t),\hat\psi^\dagger(t^\prime)\}\rangle
\,,\quad
 \imath S^{\rm a}(t;t^\prime) = -\theta(t^\prime-t)
               \langle\{\hat\psi(t),\hat\psi^\dagger(t^\prime)\}\rangle
\,.
\label{retarded and advanced Green function}
\end{equation}
 The statistical and causal (spectral) two point functions are defined
as
\begin{align}
 F_\psi(t;t^\prime) &= \frac12\left(\imath S^{-+}(t,t^\prime)
                                   + \imath S^{+-}(t,t^\prime)\right)
   = \frac{1}{2}\langle[\hat\psi(t),\hat\psi^\dagger(t^\prime)]\rangle
\nonumber\\
 \rho_\psi(t;t^\prime) &= \frac{1}{2\imath}S^c(t;t^\prime)
                        = \frac12\left(\imath S^{-+}(t,t^\prime)
                                   - \imath S^{+-}(t,t^\prime)\right)
   = \frac{1}{2}\langle\{\hat\psi(t),\hat\psi^\dagger(t^\prime)\}\rangle
 \,.
 \label{statistical and causal correlator}
\end{align}
such that $\rho_\psi(t;t)=1/2$ ($S^c(t;t)=\imath$).
By making use of the identity,
$\hat \psi^\dagger\hat \psi = (1/2)[\hat \psi^\dagger,\hat \psi]
           +(1/2)\{\hat \psi^\dagger,\hat \psi\}$,
one can obtain a simple relation between $\bar n(t)$ and $F_\psi(t;t)$:
\begin{equation}
 \bar n(t) = \langle \hat \psi^\dagger(t) \hat\psi(t)\rangle
         = \frac12 -F_\psi(t;t) \equiv \frac{1-\Delta_\psi(t)}{2}
\label{relation particle number stat corr}
\,,
\end{equation}
where in the last step we
 defined
{\footnote{The definition~\eqref{Delta psi}
is the fermionic equivalent of the invariant (phase space area)
$\Delta$ of a bosonic Gaussian state in Eq. \eqref{app1: phase space area}
of appendix~\ref{Appendix A: Bosonic density operator and entropy},
where we present an analogous derivation of the entropy for bosons.
The form of the Gaussian invariant $\Delta_{\psi}$ for fermions
is so simple because the fermionic density operator is diagonal in
$\hat{N}\equiv \hat{\psi}^{\dagger}\hat{\psi}$, implying that the fermionic density
matrix is diagonal in the fermionic particle state basis.
\label{footnote:bosonic case}}}
\begin{equation}
   \Delta_\psi(t) \equiv 2F_\psi(t;t) = 1-2\bar n(t)= \tanh\Bigl(\frac{a}{2}\Bigr)
\label{Delta psi}
\,.
\end{equation}
Note that in Eq. \eqref{relation particle number stat corr}
the average particle number is computed using time
dependent operators $\psi(t), \psi^{\dagger}(t)$, although
$\bar{n}$ in Eq. \eqref{average particle number} was computed
using Schr\"odinger picture operators $\hat{\psi}, \hat{\psi}^{\dagger}$ and the density
operator \eqref{HO:rho:Ansatz}. The time dependence of the operators
can be absorbed under the trace into the density operator, which then takes
the form \eqref{HO:rho:Ansatz} for equal-time operators.

Eq. \eqref{relation particle number stat corr}
represents a relation between the invariant of the correlators
and the Gaussian invariant of the density matrix, which are
in this single fermion case simply $F_\psi(t;t)$ and $a(t)$, respectively.
In different systems with multiple correlators and a more complicated Gaussian
density matrix such a relation can still be found. An example is the
bosonic case, discussed in footnote~\ref{footnote:bosonic case}
and in appendix~\ref{Appendix A: Bosonic density operator and entropy}, or the fermionic
field theoretical case, discussed in section
\ref{sec:Entropy generation in a fermionic quantum field theory}.\\
Note that $\bar n\in [0,1]$ and $\Delta_\psi\in [-1,1]$,
which can be appreciated from Eq. \eqref{Delta psi},
making the interpretation of $\Delta_\psi$
as the invariant phase space area of the state for fermions
dubious. It is hence better to think about $\Delta_\psi$ as
the Gaussian invariant of a fermionic state, while
$\bar n=(1-\Delta_\psi)/2\in [0,1]$ is more like the phase space area.

For thermal states, for which $\bar n_{\rm th}\in [0,1/2]$,
$\Delta_{\psi\rm th}$ acquires natural values,
$\Delta_{\psi\rm th}\in [0,1]$, and hence there is no problem.
In fact, $\Delta_{\psi}$ becomes negative only when higher energy states
are overpopulated, {\it i.e.} when $\bar n>1/2$.
Relation~\eqref{Delta psi} allows us to relate the fermionic
entropy~\eqref{HO:entropy} to the Gaussian invariant $\Delta_\psi(t)$,
\begin{equation}
 S_\psi = -\frac{1+\Delta_\psi}{2}\ln\Big(\frac{1+\Delta_\psi}{2}\Big)
      -\frac{1-\Delta_\psi}{2}\ln\Big(\frac{1-\Delta_\psi}{2}\Big)
\,,
\label{HO:entropy:2}
\end{equation}
which is to be compared with the analogous expression for bosons
in Eq. \eqref{app1:bosonic entropy} of
appendix~\ref{Appendix A: Bosonic density operator and entropy}.

\subsection{Coherent states}
\label{Coherent states}

 In order to make a connection to the existing literature
\cite{Floreanini:1987gr,Kiefer:1993fw,Benatti:2000wu,blaizot1986quantum},
here we rephrase our results in terms of fermionic coherent states
$|\theta\rangle$, defined by,
\begin{equation}
\hat\psi |\theta\rangle=\theta|\theta\rangle
\,.
\label{FHO:coherent state}
\end{equation}
When expressed in terms of Fock (particle number)
states~\eqref{Fock basis}, the coherent {\it ket} and {\it bra} states are
given by,
\begin{equation}
  |\theta\rangle = |0\rangle - \theta |1\rangle
\,,\qquad
  \langle\theta| = \langle 0| -  \langle 1| \bar\theta
\,,
\label{FHO:coherent state:2}
\end{equation}
where $\bar\theta = \theta^*$ and
Grassmann variables obey a Grassmann algebra,
$\theta_i\theta_j=-\theta_j\theta_i$ (recall that complex
conjugation for Grassmann variables is reminiscent of a hermitian conjugation,
$(\theta_i\theta_j)^* = \theta_j^*\theta_i^*=-\theta_i^*\theta_j^*$).
By making use of the well known relations,
\begin{equation*}
 \hat\psi|0\rangle = 0
\,,\qquad
 \hat\psi^\dagger|0\rangle = |1\rangle
\,,\qquad
 \hat\psi|1\rangle = |0\rangle
\,,\qquad
 \hat\psi^\dagger|1\rangle = 0
\,,
\end{equation*}
one sees that~\eqref{FHO:coherent state:2} is indeed
an eigenstate of the operator $\hat\psi$ with the eigenvalue $\theta$.
Note that the Fock space element $|0\rangle$ commutes with Grassmann variables,
while $|1\rangle=\hat\psi^{\dagger}|0\rangle$ anticommutes, such that the coherent states
$|\theta\rangle$ commute with Grassmann variables.
The coherent state $|\theta\rangle$ is fixed uniquely by the requirement
\eqref{FHO:coherent state} up to a normalisation constant
$N=1+b\bar\theta\theta$, where $b$ is a complex number.
Our choice of normalisation corresponds to
\begin{equation}
  \langle\theta|\theta\rangle = 1+ \bar\theta\theta
\,,
\label{Fcoherent state:normalisation}
\end{equation}
which is Grassmann valued.
One may attempt to normalise to unity by choosing $N=1-\bar\theta\theta/2$.
The problem with this is that then the operator $\hat\psi^\dagger$ does not
act on $|\theta\rangle$ in a desired manner. In fact, one can show that
the normalisation~\eqref{Fcoherent state:normalisation} is uniquely fixed by
the requirement,
\begin{equation}
\hat\psi^\dagger|\theta\rangle =-\frac{d}{d\theta}|\theta\rangle
\,.
\label{psi dagger on coherent state}
\end{equation}
Indeed, when $\hat\psi^\dagger$ acts on $|\theta\rangle$ defined
in~\eqref{FHO:coherent state:2} one gets $|1\rangle$, and when
the derivative $- d/d\theta$ acts on the same state, one again gets
$|1\rangle$. A different normalisation would not give this result.
Finally, as a final check of consistency we consider how the anticommutator
acts on $|\theta\rangle$,
\begin{equation*}
\{\hat\psi,\hat\psi^\dagger\}|\theta\rangle
 = \Big(-\hat\psi\frac{d}{d\theta}+\hat\psi^\dagger\theta\Big)|\theta\rangle
 = \Big(\frac{d}{d\theta}\hat\psi-\theta\hat\psi^\dagger\Big)|\theta\rangle
 = \Big(\frac{d}{d\theta}\theta+\theta\frac{d}{d\theta}\Big)|\theta\rangle
 = |\theta\rangle
\,,
\end{equation*}
as it should be from~\eqref{canonical anticommutation}.
When projected on a coherent state basis, the elements of the density
matrix~\eqref{HO:rho:solution:2} become of the form,
\begin{equation}
 \rho(\bar\theta^\prime,\theta;t)
       \equiv \langle\theta^\prime|\hat \rho(t)|\theta\rangle
    = (1-\bar n)+\bar n\bar\theta^\prime\theta
    = (1-\bar n)\exp\left(\frac{\bar n}{1-\bar n}\bar\theta' \theta\right)
\,,
\label{HO:rho:solution:3}
\end{equation}
which is not diagonal. This is to be contrasted with a diagonal {\it Ansatz}
used {\it e.g.} in Ref.~\cite{Benatti:2000wu}.

\medskip

 Let us now consider properties of the coherent state basis in more detail.
 Taking a trace of the density operator in the coherent state representation
yields,\footnote{The expression for the trace in Eq. \eqref{coherent basis: trace} can be
derived as follows: the trace of an operator $\mathcal{O}$ is in the Fock basis defined as
$\rm{Tr}[\mathcal{O}]=\langle 0|\mathcal{O}|0\rangle + \langle 1|\mathcal{O}|1\rangle$, where
the Fock space elements can be expressed in terms of coherent states as
\begin{equation*}
|0\rangle=\int d\theta \theta |\theta \rangle,\qquad |1\rangle=-\int d\theta |\theta \rangle,\qquad
\langle 0 |=\int d\bar\theta \bar\theta \langle \theta |
\,,\qquad \langle 1 |=\int d\bar\theta \langle\theta|.
\end{equation*}
}
\begin{equation}
{\rm Tr}\left[\hat\rho\right]
   = \int d\theta \int d\bar\theta \exp(\bar\theta\theta)
          \langle\theta|\hat\rho|\theta\rangle
   = \int d\theta \int d\bar\theta (1+\bar\theta\theta)
      \left[(1-\bar n)+\bar n\bar\theta\theta\right]
   = 1
\,,
\label{coherent basis: trace}
\end{equation}
where in the last step we used the usual integration rules,
$\int d\theta  = 0$,  $\int d\theta \theta = 1$.
The integration measure factor $\exp(\bar\theta\theta)$
in~\eqref{coherent basis: trace} is necessary to get the traces correctly.

 One can now use a decomposition of unity~\footnote{Note that
${\rm Tr}[I_\theta]=2$, as it should be, where $I_\theta$ is given
in Eq.~\eqref{decomposition of unity}.},
\begin{equation}
 I_\theta = \int d\bar\theta \int d\theta
          \exp(-\bar\theta\theta)|\theta\rangle\langle\theta|
\,,
\label{decomposition of unity}
\end{equation}
to recast $\hat\rho$ as,
\begin{align}
\hat\rho &= \int d\bar\theta^\prime d\theta^\prime
                            {\rm e}^{-\bar\theta^\prime\theta^\prime}
              |\theta^\prime\rangle\langle\theta^\prime|\hat\rho
                 \int d\bar\theta d\theta{\rm e}^{-\bar\theta\theta}
|\theta\rangle\langle\theta|
  = \int d\bar\theta^\prime d\theta^\prime
                    {\rm e}^{-\bar\theta^\prime\theta^\prime}
              |\theta^\prime\rangle
             \left[(1-\bar n)+\bar n\bar\theta^\prime\theta\right]
               \int d\bar\theta d\theta
                              {\rm e}^{-\bar\theta\theta} \langle\theta|
\nonumber\\
 &\equiv \int d\bar\theta^\prime d\theta^\prime
                            {\rm e}^{-\bar\theta^\prime\theta^\prime}
               \int d\bar\theta d\theta
                              {\rm e}^{-\bar\theta\theta}
    \hat\rho(\bar\theta^\prime,\theta,t)
\,,
\label{rho in coherent state rep}
\end{align}
where $\hat\rho(\bar\theta^\prime,\theta,t)$ are
 elements of the density operator in the coherent state representation
 (see Eq. \eqref{HO:rho:solution:3}),
\begin{equation}
\hat\rho(\bar\theta^\prime,\theta;t)
   =  |\theta^\prime\rangle
       \left[(1-\bar n)
             +\bar n\bar\theta^\prime\theta\right]
       \langle\theta|
   =  |\theta^\prime\rangle
       \rho(\bar\theta^\prime,\theta;t)
       \langle\theta|
\,,
\label{rho in coherent state rep:2}
\end{equation}
with $\rho(\bar\theta^\prime,\theta;t)=Z^{-1}\exp(M\bar{\theta}'\theta)$
given in Eq.~\eqref{HO:rho:solution:3}.
$\hat\rho(\bar\theta^\prime,\theta;t)$
of Eq. \eqref{rho in coherent state rep:2} is obviously non-diagonal.
However $\hat{\rho}$ of Eq. \eqref{rho in coherent state rep}
can be cast in a diagonal basis by inserting the {\it Ansatz}
\begin{equation}
\rho(\theta,\bar\theta^\prime;t)=\int d\bar{\zeta}d\zeta
{\rm e}^{\bar{\theta}'\zeta + \bar{\zeta}\theta} P(\zeta)
\,,
\label{ansatz P representation}
\end{equation}
such that
\begin{equation}
\hat{\rho}=\int d\bar{\zeta}d\zeta
|\zeta\rangle P(\zeta) \langle \zeta |
\,.
\label{P representation}
\end{equation}
This is the so-called Glauber $P$ representation
\cite{Glauber:1963tx} for fermions.
Inverting \eqref{ansatz P representation}
the function $P(\zeta)$ is related to the density
matrix in the diagonal elements
of the coherent state basis as
\begin{equation}
P(\zeta)=\int d\theta d\bar{\theta} {\rm e}^{-\bar{\theta}\zeta-\bar{\zeta}\theta}
\langle \theta | \hat{\rho} |\theta \rangle
\,.
\label{relation P-rep diagonal coherent}
\end{equation}
The elements of the density matrix in the coherent state basis
have been found in \eqref{HO:rho:solution:3},
and by integrating over $\theta,{\bar\theta}$ in
\eqref{relation P-rep diagonal coherent} one finds:
\begin{equation}
P(\zeta)=\bar{n}+(1-\bar{n})\zeta\bar{\zeta}
\,.
\label{Pzeta in our model}
\end{equation}
It is possible to return to the Fock basis
via the $P$ representation \eqref{P representation}
using Eqs. \eqref{FHO:coherent state:2} and \eqref{Pzeta in our model}
and integrating over $\bar{\zeta},\zeta$,
\begin{equation}
 \hat\rho(t) =  \sum_{n=0}^1|n\rangle[(1-\bar n)+(2\bar n-1)n]\langle n|
             = |0\rangle(1-\bar n)\langle 0|
               + |1\rangle\bar n\langle 1|
\,.
\label{rho in particle number rep}
\end{equation}
For one degree of freedom the (diagonal)
Fock number basis is by far superior to
the coherent state basis~\eqref{P representation} for
studying properties of the fermionic density operator, an important example
being the von Neumann entropy defined in~\eqref{entropy:vN} and
calculated in~\eqref{HO:entropy}. The reason is that the Fock states
are orthogonal, contrary to the coherent states
which satisfy $\langle \zeta | \theta \rangle={\rm e}^{\bar{\zeta}\theta}$,
see Eq. \eqref{Fcoherent state:normalisation}.
Still, the von Neumann entropy can be derived from
the density operator in the coherent state basis
by using the replica trick, which is demonstrated
in appendix \ref{Appendix D: Entropy via the replica trick in coherent state basis}.
There we also generalise to $N$ fermionic degrees of freedom,
a case which is discussed in more detail in section~\ref{sec: Ndegrees of freedom}.

\subsection{Fermionic interactions in quantum mechanics}
\label{sec:Fermionic dynamics}

Interactions can be included in the quantum mechanical
fermionic theory \eqref{HO:lagrangian} by introducing
general current terms into the Lagrangian
\begin{equation}
 L_\psi \equiv L_0 +L_{\rm int}
\,;\qquad
L_{\rm int} =  - \hat j_\psi^\dagger\hat\psi- \hat\psi^\dagger\hat j_\psi
\,.
\label{current:lagrangian}
\end{equation}
Formally the linear current terms can be absorbed into
the free field theory \eqref{HO:lagrangian}
by shifting the fermionic fields, see
appendix~\ref{Appendix B: Fermionic shift and diagonalisation}.
For these free shifted fermionic fields the von Neumann entropy is conserved.

In realistic situations it is very hard to have a complete information about
the current operator $\hat j_\psi(t)$ however, which makes the diagonalisation
procedure~\eqref{shifted field} impracticable, or even impossible.
Namely, in condensed matter systems, the coupling current is often
given by a superposition of many (fermionic) degrees of freedom, whose
precise time evolution is not known. In a quantum field theoretic setting
one can have for example a Yukawa coupling term,
$-y\hat\phi(x)\hat{\bar\psi}(x)\hat\psi(x)$, such that
the coupling current $\hat j_\psi$ corresponds to a composite operator,
$\hat j_\psi(x) = y\hat\phi(x)\hat{\bar\psi}(x)$, making the diagonalisation
procedure~\eqref{shifted field} very hard, if not impossible.
 For that reason we adopt here the point of view that no (useful) information
is known about the evolution of the current ${\hat j}_\psi$.
This {\it loss of information} leads to entropy generation,
which is what we study next.

The simplest nontrivial example
is the quantum mechanical case when the current consists of $N$ environmental
oscillators in thermal equilibrium. In this case,
\begin{equation}
 \hat j_\psi(t)= \sum_{i=1}^N\lambda_i\hat\psi_{q_i}(t)
\,,
\label{current:N oscillators}
\end{equation}
where the $\hat \psi_{q_i}$ represent the environmental fermionic oscillators.
The form of the current~(\ref{current:N oscillators}) is motivated by
mass mixing, which can be considered as an approximation to
the Yukawa coupling, {\it cf.} Eqs.~(\ref{Yukawa}--\ref{Yukawa:gaussian}).
The system is represented by a single fermionic oscillator $\psi_x$ which is
coupled bilinearly to the environmental oscillators through couplings $\lambda_i$.
The interaction between the environmental oscillators is assumed to be zero
in our toy model. The {\it loss of information} in this case is that we cannot
observe (correlations of) the environmental oscillators, nor its interaction with
the system.
The complete action of system, environment and interactions in our toy model is
\begin{equation}
S[\hat{\psi}_x,\{\hat{\psi}_{q_i}\}]
=\int dt \left\{ L_{\text{S}}[\hat{\psi}_x]
+ L_{\text{E}}[\{\hat{\psi}_{q_i}\}]
+ L_{\text{int}}[\hat{\psi}_x,\{\hat{\psi}_{q_i}\}]\right\}
\,,
\label{BCF:bilinearfermionmodel}
\end{equation}
with
\begin{align}
\nonumber L_{\text{S}}[\hat{\psi}_x]&=
\hat{\psi}_x^{\dagger} (\imath\partial_t-\omega_0)\hat{\psi}_x\\
\nonumber L_{\text{E}}[\{\hat{\psi}_{q_i}\}]&=
\sum_{i=1}^N\hat{\psi}_{q_i}^{\dagger} (\imath\partial_t-\omega_i)\hat{\psi}_{q_i}\\
L_{\text{int}}[\hat{\psi}_x,\{\hat{\psi}_{q_i}\}]&=
-\sum_{i=1}^N\lambda_i\left(\hat{\psi}_x^{\dagger}\hat{\psi}_{q_i}
+\hat{\psi}_{q_i}^{\dagger}\hat{\psi}_x\right)
\,.
\end{align}
Note that by the hermiticity of $L_{\rm int}$,
all $\lambda_i^*=\lambda_i$ are real.
The fermionic oscillators only depend on time,
\textit{i.e.} $\hat{\psi}_x=\hat{\psi}_x(t)$ and
$\hat{\psi}_{q_i}=\hat{\psi}_{q_i}(t)$. The anticommutation relations
satisfied by $\hat{\psi}_x$ and $\hat{\psi}_{q_i}$ are
\begin{align}
\nonumber \{\hat{\psi}_x(t),\hat{\psi}_x^{\dagger}(t)\}&=1\\
\nonumber \{\hat{\psi}_{q_i}(t),\hat{\psi}_{q_j}^{\dagger}(t)\}&=\delta_{ij}
\,,
\end{align}
with all others being zero.
Of our interest are the statistical correlators
\eqref{statistical and causal correlator},
which are for our model defined as
\begin{align}
\nonumber F_{xx}(t;t')&=\frac12 \langle [ \hat{\psi}_x(t),\hat{\psi}_x^{\dagger}(t')]\rangle \\
\nonumber F_{q_iq_j}(t;t')&=\frac12 \langle [ \hat{\psi}_{q_i}(t),\hat{\psi}_{q_j}^{\dagger}(t')]\rangle\\
\nonumber F_{xq_i}(t;t')&=\frac12 \langle [ \hat{\psi}_{x}(t),\hat{\psi}_{q_i}^{\dagger}(t')]\rangle\\
F_{q_ix}(t;t')&=\frac12 \langle [ \hat{\psi}_{q_i}(t),\hat{\psi}_{x}^{\dagger}(t')]\rangle
\,,
\label{BCF:statCorrenvsystem}
\end{align}
where $\hat{\psi}_x(t)$ and $ \hat{\psi}_{q_i}(t)$ are here Heisenberg picture operators.
Our goal is to calculate the entropy for the system. For a free
fermionic theory the entropy is given by Eqs. \eqref{Delta psi}
and \eqref{HO:entropy:2}. Without interactions,
the Gaussian invariant $\Delta$ is constant and the entropy is conserved.
If we switch on interactions the Gaussian invariant and the entropy change
in time. The entropy of the system with the Gaussian
{\it Ansatz}~\eqref{HO:rho:Ansatz}
for $\hat{\rho}$ with time dependent $a(t)$ is related to $\Delta_{xx}$
as in Eq. \eqref{HO:entropy:2}:
\begin{equation}
S_x(t)=-\frac{1+\Delta_{xx}(t)}{2}\ln\left(\frac{1+\Delta_{xx}(t)}{2}\right)
-\frac{1-\Delta_{xx}(t)}{2}\ln\left(\frac{1-\Delta_{xx}(t)}{2}\right)
\,,
\label{BCF:entropysytem}
\end{equation}
with
\begin{equation}
\Delta_{xx}(t)=2F_{xx}(t;t)\equiv 1-2\bar{n}_{xx}(t),\label{BCF:invareafermionsystem}
\end{equation}
and $\bar{n}_{xx}(t)$ the average particle number for the system fermions.
The proper way to derive the entropy of the system is to
trace over the environmental degrees
of freedom in the density operator, and calculate the entropy
from this \textit{reduced} density operator. The
corresponding reduced von Neumann entropy is the same as Eq.
\eqref{BCF:entropysytem},
\textit{i.e.} $S^{\rm red}_{\rm vN}(t)=S_x(t)$.\footnote{The proof goes
as follows. The reduced density matrix is defined
as $\hat{\rho}^{\rm red}={\rm Tr}_{\rm E}[\hat{\rho}]$, where the subscript
${\rm E}$
denotes the environment, which in this example is the group of
oscillators $\{\psi_{q_i}\}$. The reduced von Neumann entropy is the
usual $S^{\rm red}_{\rm vN}(t)=-{\rm Tr}[\hat{\rho}^{\rm red}\ln\hat{\rho}^{\rm red}]$.
Now most importantly, if we calculate correlators of the system, we have
\begin{equation}
(1-\Delta_{xx})/2=\langle \hat{\psi}^{\dagger}_x\hat{\psi}_x\rangle
= {\rm Tr}[\hat{\rho}\hat{\psi}^{\dagger}_x\hat{\psi}_x]
= \sum_{n_x,n_{q_i}} \langle n_x|\langle n_{q_1}|..\langle n_{q_N}|
\hat{\rho}\hat{\psi}^{\dagger}_x\hat{\psi}_x |n_{q_N}\rangle..|n_{q_1}\rangle |n_{x}\rangle
= {\rm Tr}_{\rm S}[\hat{\rho}^{\rm red}\hat{\psi}^{\dagger}_x\hat{\psi}_x].
\label{proof reduced density matrix}
\end{equation}
Here $\hat{\psi}_x, \hat{\psi}^{\dagger}_x$ and $\hat{\rho}$ are taken to be
in the Schr\"odinger picture, which ensures that $\hat{\psi}^{\dagger}_x\hat{\psi}_x$
does not evolve with time, whereas $\hat{\rho}$ controls the evolution. The trace
is taken over a complete set of orthogonal, time independent Fock states.
Eq. \eqref{proof reduced density matrix} shows that the correlators,
and thus the invariant area \eqref{BCF:invareafermionsystem}
and entropy \eqref{BCF:entropysytem} for the \textit{system} are
the same whether you first trace over the environment in the density matrix
or you consider the full density matrix.
}
Thus, in order to investigate the growth of entropy for the system fermionic oscillator $\hat{\psi}_x$,
we should find the statistical correlator $F_{xx}(t,t)$ defined in Eq. \eqref{BCF:statCorrenvsystem}.
The equations of motion for the fermionic operators follow from the action \eqref{BCF:bilinearfermionmodel},
\begin{align}
\nonumber (\imath\partial_t-\omega_0)\hat{\psi}_x(t)&=\sum_{i=1}^N\lambda_i\hat{\psi}_{q_i}(t)\\
\nonumber(\imath\partial_t-\omega_i)\hat{\psi}_{q_i}(t)&=\lambda_i\hat{\psi}_{x}(t)\\
\nonumber(-\imath\partial_t-\omega_0)\hat{\psi}^{\dagger}_x(t)&=\sum_{i=1}^N\lambda_i\hat{\psi}^{\dagger}_{q_i}(t)\\
(-\imath\partial_t-\omega_i)\hat{\psi}^{\dagger}_{q_i}(t)&=\lambda_i\hat{\psi}^{\dagger}_{x}(t)
\,.
\label{BCF:eompsix}
\end{align}
 From these equations of motion we can derive coupled differential equations for the statistical equal-time correlators
\begin{align}
\nonumber \imath\partial_t F_{xx}(t;t)
&=\sum_{i=1}^N\lambda_i\left(F_{q_ix}(t;t)-F_{xq_i}(t;t)\right)\\
\nonumber (\imath\partial_t-(\omega_0-\omega_j))F_{xq_j}(t;t)
&= -\lambda_jF_{xx}(t;t)+\sum_{i=1}^N\lambda_iF_{q_iq_j}(t;t)\\
\nonumber (\imath\partial_t-(\omega_j-\omega_0))F_{q_jx}(t;t)
&=\lambda_jF_{xx}(t;t)-\sum_{i=1}^N\lambda_iF_{q_jq_i}(t;t)\\
(\imath\partial_t-(\omega_i-\omega_j))F_{q_iq_j}(t;t)
&= \lambda_iF_{xq_j}(t;t)-\lambda_jF_{q_ix}(t;t)
\,.
\label{BCF:eomFqq}
\end{align}
These conditions can be solved with suitable initial conditions.
We take the system fermionic oscillator
to be initially in a state with average particle number zero.
The environmental fermionic oscillators are assumed to be
in a thermal state according to the Fermi-Dirac distribution
with energy $E_i=\omega_i$, see Eq. \eqref{Fermi Dirac}.
Thus
\begin{align}
\nonumber F_{xx}(t_0;t_0)&=\frac{1}{2}\\
\nonumber F_{q_iq_j}(t_0;t_0)&=\delta_{ij}\frac{1}{2}\tanh\left(\frac{\beta\omega_i}{2}\right)\\
F_{xq_i}(t_0;t_0)&=F_{q_ix}(t_0;t_0)=0
\,.
\label{BCF:initialFxq}
\end{align}
With these initial conditions Eqs. \eqref{BCF:eomFqq} can be solved.
We first treat the simple case of two coupled oscillators,
then the general case of $N$ coupled oscillators.

\subsubsection{Two coupled fermionic oscillators}

\begin{figure}[t!]
    \begin{minipage}[t]{.45\textwidth}
        \begin{center}
\includegraphics[width=\textwidth]{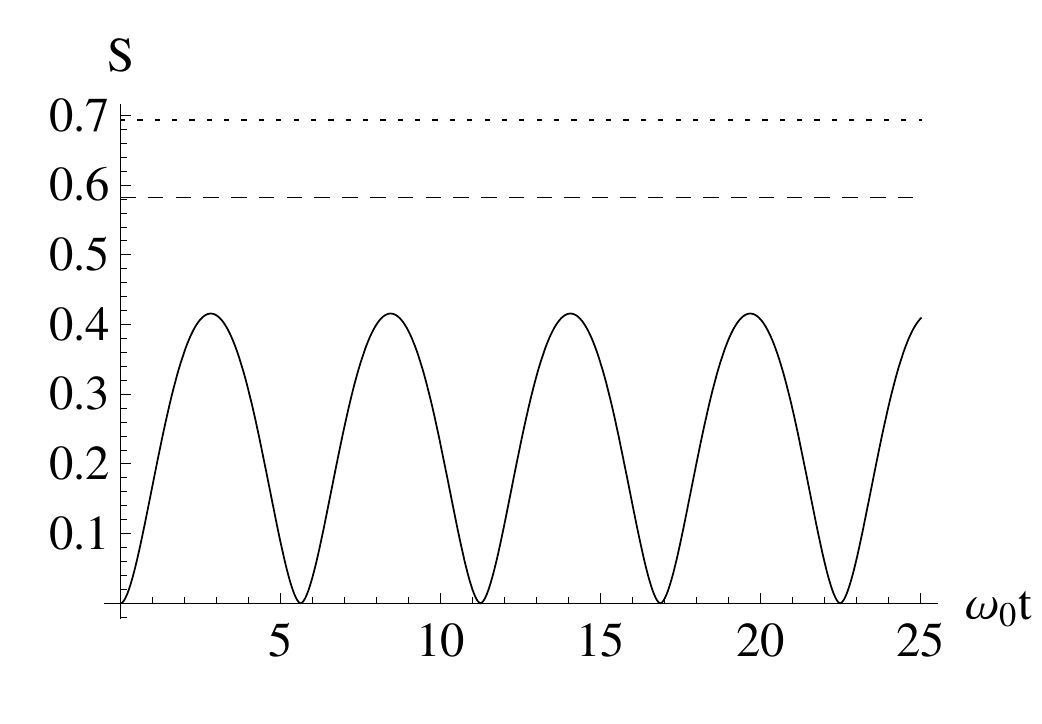}
   {\em \caption{System entropy as a function of $\omega_0 t$ for $N=1$ environmental oscillator.
   The parameters are $\omega_1=1.5\omega_0$, $\lambda=0.5\omega_0$ and $\beta=(\omega_0)^{-1}$.
   The entropy oscillates between zero and the thermal entropy \eqref{thermal entropy} (dashed line).
   The maximum entropy $S_{\text{max}}=\ln(2)$ is indicated with the dotted line.
   \label{fig:2coupledfermions1} }}
        \end{center}
    \end{minipage}
\hfill
    \begin{minipage}[t]{.45\textwidth}
        \begin{center}
\includegraphics[width=\textwidth]{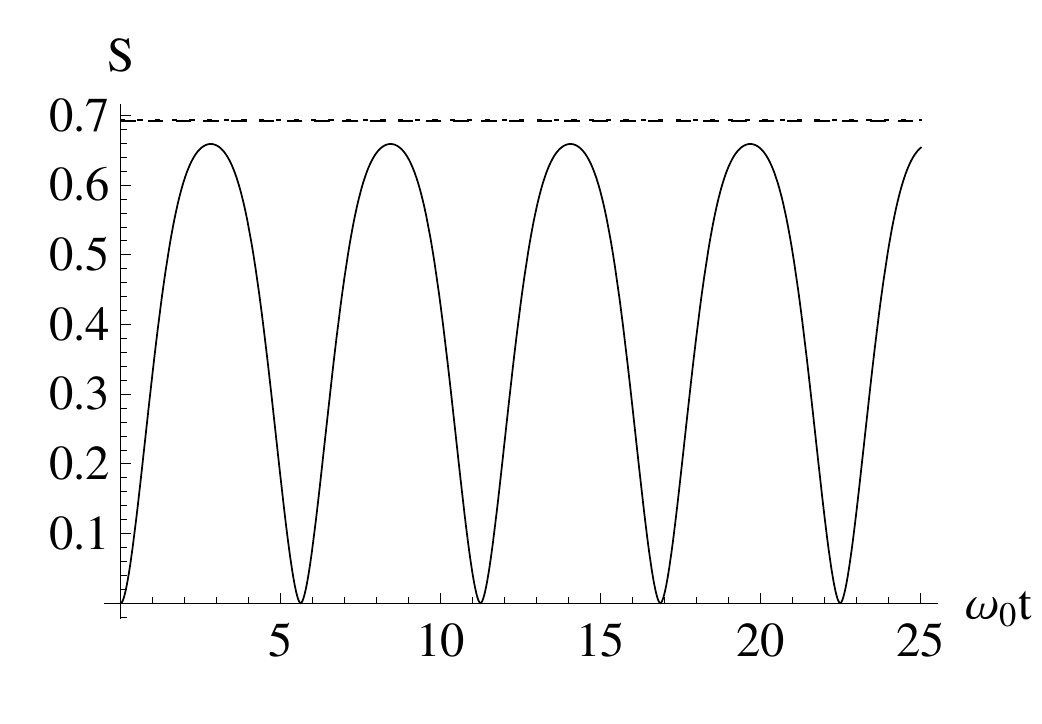}
   {\em \caption{System entropy as a function of $\omega_0 t$ for $N=1$ environmental oscillator.
   The parameters are $\omega_1=1.5\omega_0$, $\lambda=0.5\omega_0$ and $\beta=0.1(\omega_0)^{-1}$.
   In this case the entropy almost reaches the thermal entropy \eqref{thermal entropy} (dashed line),
   which is in turn almost equal to the maximum value $S_{\text{max}}=\ln(2)$ (dotted line).
   \label{fig:2coupledfermions2} }}
        \end{center}
    \end{minipage}
\end{figure}

For two coupled fermionic oscillators we consider the case $N=1$
in the action \eqref{BCF:bilinearfermionmodel}. Thus there is one
environmental oscillator $\hat{\psi}_q\equiv \hat{\psi}_{q_1}$
coupled to the system oscillator $\hat{\psi}_x$ through a
coupling $\lambda\equiv \lambda_1$. This simple example can be
solved analytically. The procedure is explained in
appendix~\ref{Appendix C: Exact entropy for two coupled fermions}.
As a final result we find an explicit expression for the Gaussian invariant of
the system represented by the fermions $\hat{\psi}_x$,
\begin{equation}
\Delta_{xx}(t)
=1-2\bar{n}_{\text{E}}\left(\frac{2\lambda}{\bar{\omega}}\right)^2
                \sin^2\Big[\frac{\bar{\omega}}{2}(t-t_0)\Big]
\,,
\label{BCF:invareatwofermions}
\end{equation}
where
\begin{align}
\nonumber \bar{n}_{\text{E}}&=\frac{1}{{\rm e}^{\beta\omega_1}+1}\\
\bar{\omega} &= \sqrt{(\omega_0-\omega_1)^2+4\lambda^2}
\,,
\end{align}
with the frequencies $\omega_0$ and $\omega_1$ of the system and environment
oscillators, respectively.
The entropy of the system $S_x$ is subsequently found using Eq. \eqref{BCF:entropysytem}.
Figures \ref{fig:2coupledfermions1} and \ref{fig:2coupledfermions2}
show the evolution of entropy for a system coupled to
one environmental oscillator with $\omega_1=1.5\omega_0$
and $\lambda=0.5\omega_0$ at different values of $\beta$.
The dashed line indicates the entropy in the case
when the system is completely thermalised \eqref{thermal entropy}
and the dotted line is the maximum entropy $S_{\text{max}}=\ln(2)$.
The Gaussian invariant~\eqref{BCF:invareatwofermions}
satisfies the correct properties: initially $\Delta_{xx}(t_0)=1$
and the entropy \eqref{BCF:entropysytem} is zero. For zero coupling ($\lambda=0$),
the two oscillators do not interact and the Gaussian invariant remains conserved,
leaving zero entropy. For general coupling the Gaussian invariant oscillates
between 1 and some value $>0$ with an angular frequency $\bar{\omega}$.
The corresponding entropy then oscillates between $0$ and the thermal entropy.
Only in the limit when $\beta\rightarrow 0$ and $\omega_0\rightarrow \omega_1$
(resonant regime)
the maximum entropy  $S_{\text{max}}=\ln(2)$ is reached (for $\Delta_{xx}=0$).

\subsubsection{$N+1$ coupled fermionic oscillators}
\label{N+1 coupled fermionic oscillators}

\begin{figure}[t!]
    \begin{minipage}[t]{.45\textwidth}
        \begin{center}
\includegraphics[width=\textwidth]{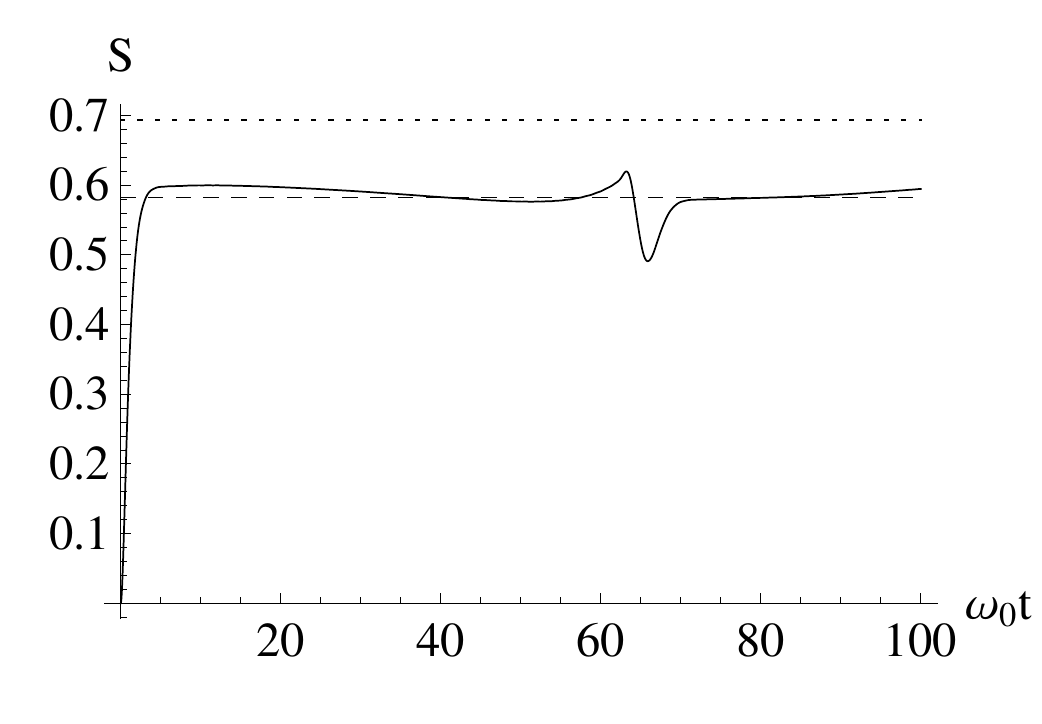}
   {\em \caption{System entropy as a function of $\omega_0 t$ for $N=50$ environmental oscillators.
   The parameters are: $\omega_i=0.1i\times\omega_0,~i=1,..,50$,
   $\lambda=0.15\omega_0$ and $\beta=(\omega_0)^{-1}$.
   The entropy rapidly reaches value of the thermal entropy $S_{\rm th}$ \eqref{thermal entropy},
   indicated by the dashed line. The dotted line is the maximum entropy $S_{\text{max}}=\ln(2)$.
   At specific $\omega_0 t$ there are small fluctuations of the system entropy.
   \label{fig:50coupledfermions5} }}
        \end{center}
    \end{minipage}
\hfill
    \begin{minipage}[t]{.45\textwidth}
        \begin{center}
\includegraphics[width=\textwidth]{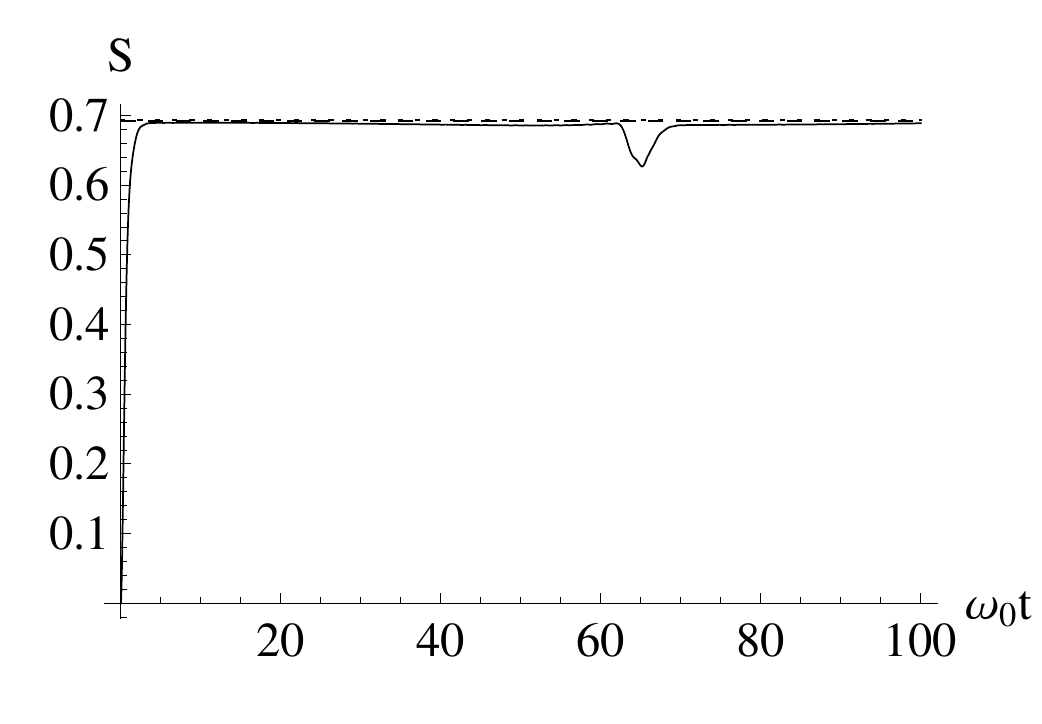}
   {\em \caption{System entropy as a function of $\omega_0 t$ for $N=50$ environmental oscillators.
   The parameters are: $\omega_i=0.1i\times \omega_0,~i=1,..,50$,
   $\lambda=0.15\omega_0$ and $\beta=0.1(\omega_0)^{-1}$.
   The entropy again rapidly reaches the thermal entropy
   $S_{\rm th}$ \eqref{thermal entropy} (dashed line), which at such high
   temperatures (low $\beta$) almost coincides with the
   maximum entropy $S_{\text{max}}=\ln(2)$ (dotted line).
   \label{fig:50coupledfermions6} }}
        \end{center}
    \end{minipage}
\end{figure}

\begin{figure}[t!]
    \begin{minipage}[t]{.45\textwidth}
        \begin{center}
\includegraphics[width=\textwidth]{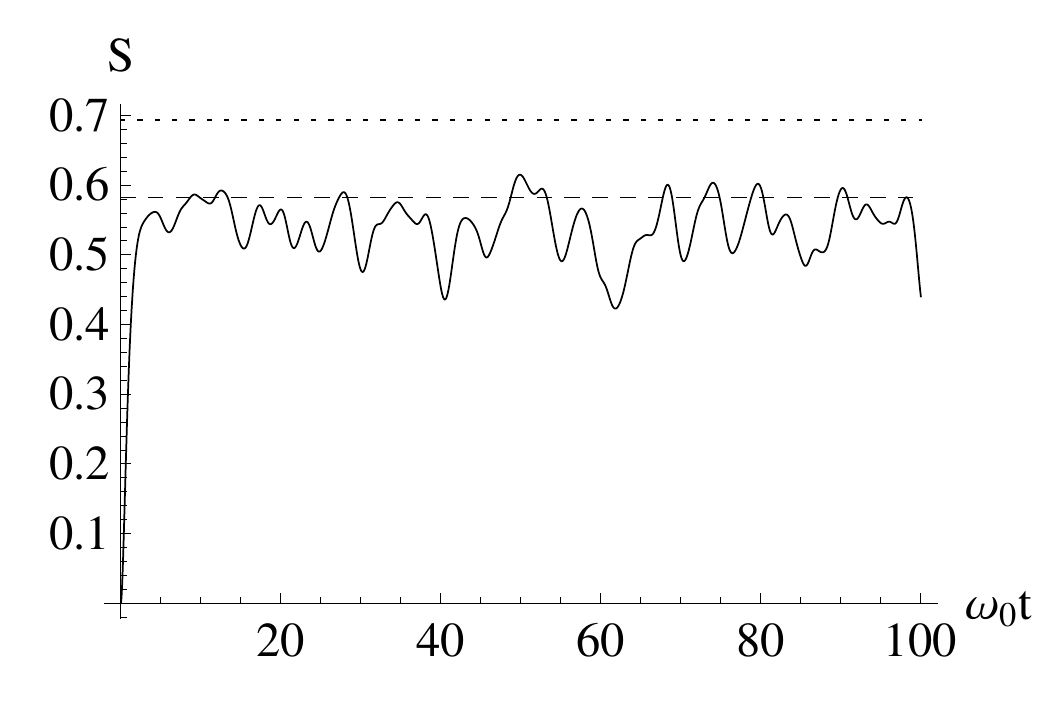}
   {\em \caption{Same plot as Fig. \ref{fig:50coupledfermions5},
   but here the environmental frequencies have been randomly selected
   in the same interval, \textit{i.e.} $\omega_i\in [0-5]\times\omega_0$.
   Due to the random choice of the environmental frequencies,
   the entropy randomly fluctuates around the thermal value.
   \label{fig:50coupledfermions3} }}
        \end{center}
    \end{minipage}
\hfill
    \begin{minipage}[t]{.45\textwidth}
        \begin{center}
\includegraphics[width=\textwidth]{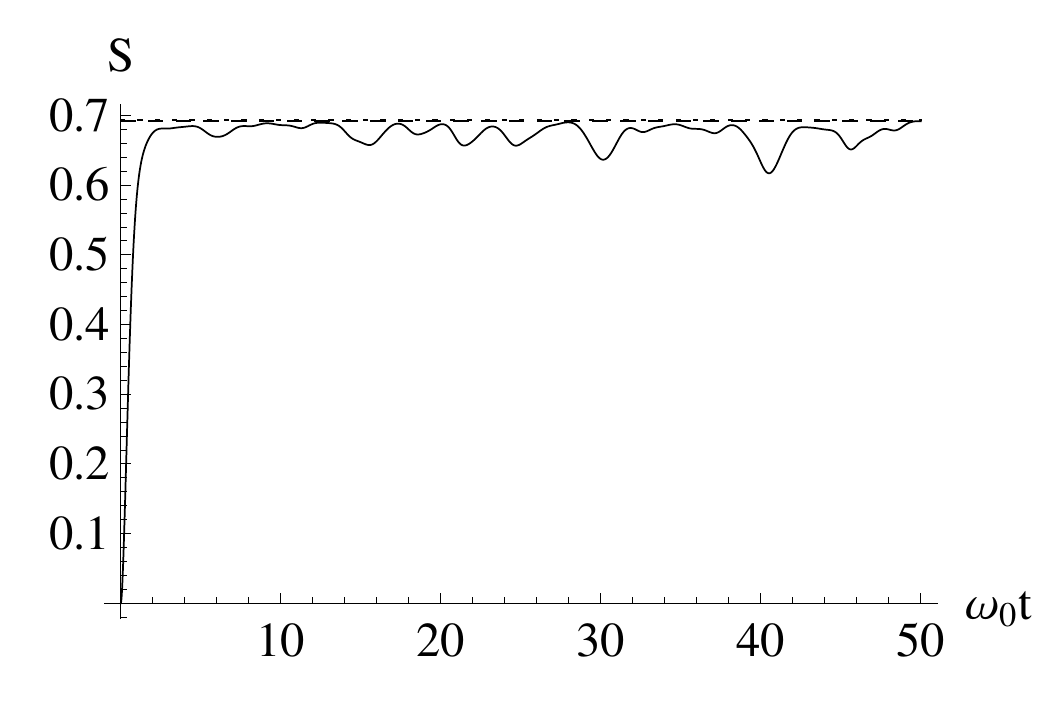}
   {\em \caption{Same plot as Fig. \ref{fig:50coupledfermions6},
   but here the environmental frequencies have been randomly selected
   in the same interval, \textit{i.e.} $\omega_i\in [0-5]\times\omega_0$.
   Again the entropy fluctuates around the thermal value much more
   frequently due to the random choice of environmental frequencies.
   \label{fig:50coupledfermions4} }}
        \end{center}
    \end{minipage}
\end{figure}

\begin{figure}[t!]
            \begin{center}
\includegraphics[width=.45\textwidth]{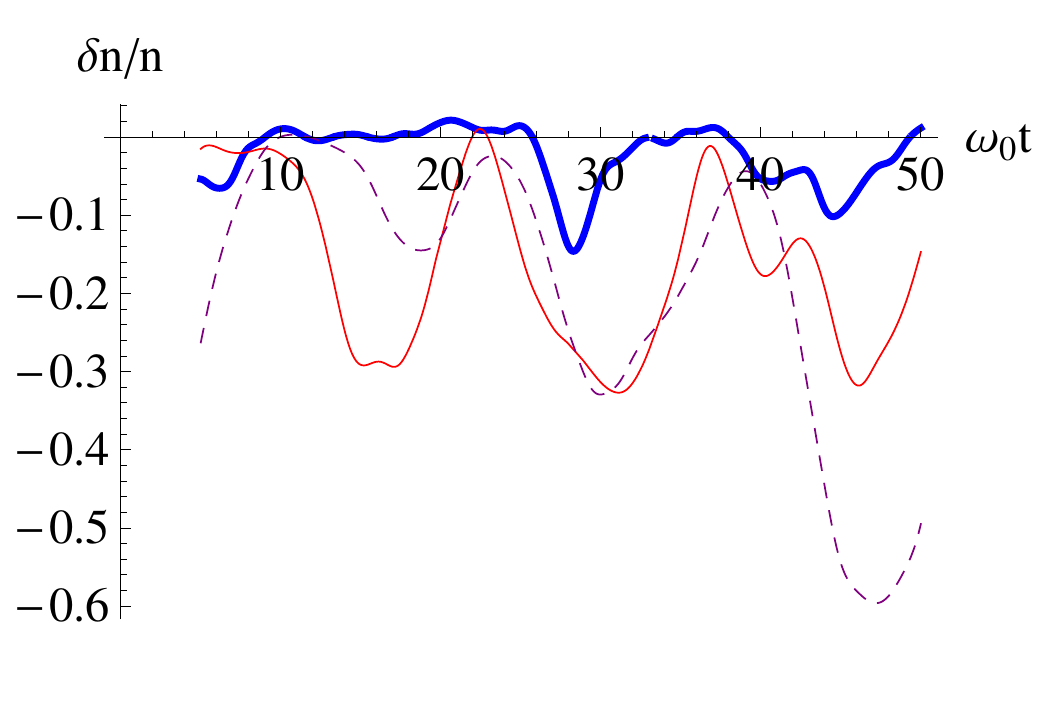}
   {\em \caption{Amplitude of fluctuations in statistical particle number
   at late time for N=25 (purple, dashed), N=50 (red, solid) and
   N=100 (blue, thick solid) environmental oscillators. $\delta n/n$ is defined as
   $\delta n/n\equiv (\bar{n}_{xx}-n_{\rm th})/n_{\rm th}$,
   where $\bar{n}_{xx}$ is the statistical particle number of the system
   \eqref{BCF:invareafermionsystem} and $n_{\rm th}=(\exp(\beta \omega_0)+1)^{-1}$.
   For all lines $\lambda=0.1$, $\beta=0.5$ and
   the environmental frequencies have been chosen randomly in the same range,
   $\omega_i\in [0-5]\times\omega_0$. In general, there are fluctuations in
   the late time entropy due to the finite amount of environmental oscillators
   that constructively and destructively interfere. The plot clearly shows that
   the amplitude of these late time fluctuations decreases as $N$ increases.
   \label{fig:flucsNdependence} }}
        \end{center}
\end{figure}

In this section we consider the more general case of one system
oscillator bilinearly coupled to $N$ environmental fermions.
In order to find the growth of entropy of the system
Eqs. \eqref{BCF:eomFqq} must be solved for the statistical correlators
with initial conditions \eqref{BCF:initialFxq}. This can be
done numerically.
We have used the $N=1$ case treated analytically above as the test case for
our numerical studies.
For simplicity the system oscillator couples
equally to all the environmental oscillators,
\textit{i.e.} $\lambda_i\equiv \lambda$.
If the frequencies of the environmental oscillators are taken
in a narrow range away from $\omega_0$,
they will effectively behave as a single oscillator, leading to
similar plots as figures \ref{fig:2coupledfermions1} and
\ref{fig:2coupledfermions2}.\\
In figures \ref{fig:50coupledfermions5}--\ref{fig:50coupledfermions4}
the system entropy is calculated by taking 50 environmental oscillators
with frequencies in the range of $[0-5]\times \omega_0$.
The equal couplings
to the system oscillator are $\lambda_i\equiv\lambda=0.15\omega_0$.
In figures \ref{fig:50coupledfermions5}--\ref{fig:50coupledfermions6}
the environmental frequencies are equally spaced, \textit{i.e.}
$\omega_i=0.1i\omega_0,~i=1,..,50$,
with $\beta=1$ and $\beta=0.1$, respectively.
The system entropy rapidly increases to the value of the entropy
in case the system is completely thermalised, see Eq. \eqref{thermal entropy}.
For higher environmental temperature (lower $\beta$)
the late time entropy gets closer to the maximum entropy $S_{\max}=\ln(2)$,
which is only reached for $\beta\rightarrow 0$.
In general, there will be fluctuations in the entropy due to
constructive and destructive interference of the environmental
oscillators. Due to the specific distribution of environmental oscillators
in Figs. \ref{fig:50coupledfermions5}--\ref{fig:50coupledfermions6},
the entropy is almost constant with small fluctuations at very regular times
when the oscillators interfere in a constructive or destructive manner.
As the number of oscillators increases these features occur on longer
timescales and the amplitude of fluctuations decreases. In the limit
when $N\rightarrow \infty$ these features disappear altogether.
\\
In figures \ref{fig:50coupledfermions3}--\ref{fig:50coupledfermions4}
the environmental frequencies of the 50 oscillators
are randomly selected in the interval $[0-5]\times \omega_0$.
The entropy increases to the same values as in figures
\ref{fig:50coupledfermions5}--\ref{fig:50coupledfermions6},
but due to the random choice of frequencies the late-time
entropy contains some random fluctuations. As $N$ increases,
the amplitude of the fluctuations $\delta n/n$ decreases.
In figure
\ref{fig:flucsNdependence} we have shown the amplitude of
fluctuations in late time statistical particle number
for $N=25$, $N=50$ and $N=100$ and it demonstrates that
the size of these fluctuations becomes smaller as $N$
becomes larger, and it is consistent with the expected behaviour,
$\delta n/n\sim 1/\sqrt{N}$. For infinitely many coupled environmental
oscillators the statistical particle number, or entropy,
at late time becomes more and more stable.

 We have also studied the growth of system's entropy for bilinearly
coupled environmental oscillators. For a finite number of oscillators
the fluctuations in particle number and entropy are large,
making it difficult to extract a specific growth rate, in particular
because in this case the system does not seem to approach exponentially
its final statistical occupation number.
For $\lambda N\gg 1$, when oscillations are small
and a relatively stable late time entropy is reached,
a rough estimate of the growth rate can be made. The rate
can be defined as $\Gamma=\dot{\bar{n}}(t)/(n_{\rm th}-\bar{n}(t))$,
and can be estimated as $\Gamma \sim \lambda \omega_0$,
plus an additional weak dependence on temperature. The growth rate
can be studied more accurately in true interacting
quantum field theories. In that case one system field mode interacts
with infinitely many environmental modes through quantum loops, such that
the thermalisation of the system field can be quite accurately described
by the perturbative rate, which in the high temperature regime is propotional to $T$.
At late times the system's entropy settles to its thermal
equilibrium value. This has been tested numerically
for bosons with a cubic interaction in Ref.~\cite{Koksma:2011dy}.
As with regards to fermions,
the quantum field theoretical description of fermionic entropy
will be discussed in the next section.


\section{Entropy generation in a fermionic quantum field theory}
\label{sec:Entropy generation in a fermionic quantum field theory}

 We shall now consider the entropy of a fermionic field theory whose
action is given by,
\begin{equation}
S[\psi] = \int d^4 x {\cal L}_\psi
 \,,\qquad
{\cal L}_\psi = \bar \psi(x)\imath\gamma^\mu\partial_\mu \psi(x)
              - m\bar \psi(x)\psi(x) + {\cal L}_{\rm int}
\,,\qquad {\cal L}_{\rm int} = - \bar j_\psi(x)\psi(x) - \bar \psi(x) j_\psi(x)
\,,
\label{QFT:action}
\end{equation}
where $\psi(x)$ is a space-time spinor, $\bar\psi(x)=\psi^\dagger\gamma^0$,
 $j_\psi(x)$ is a spinorial current, $\bar j_\psi(x)=j_\psi^\dagger\gamma^0$
and $\gamma^\mu$ are Dirac's matrices obeying a Clifford algebra with an
anticommutation relation,
\begin{equation}
 \{\gamma^\mu,\gamma^\nu\} = 2\eta^{\mu\nu}\,,\qquad
  \eta^{\mu\nu} = {\rm diag}(1,-1,-1,-1)
\,.
\label{anticommutation relation}
\end{equation}
The action \eqref{QFT:action} is a general {\it Ansatz} describing
many realistic systems. Examples include:
a fermionic field in a heat bath of many fermionic degrees of freedom
(similar to the quantum mechanical case in section~\ref{sec:Fermionic dynamics});
one quark
flavour coupled to other quark flavours through the CKM matrix
\cite{Cabibbo:1963yz,Kobayashi:1973fv};
one neutrino flavour coupled to other neutrino flavours through the PMNS matrix
\cite{Pontecorvo:1967fh,Maki:1962mu}; but also
many systems with true interactions, such as Yukawa type
$j_{\psi_i}\rightarrow y_{ij}\bar{\psi}_j\phi$
with $\phi$ a scalar field. \\
In the correlator approach to decoherence
\cite{Campo:2008ju,Campo:2008ij,
Giraud:2009tn,Koksma:2010dt,Koksma:2009wa,
Koksma:2010zi,Koksma:2011fx,Koksma:2011dy}
the Gaussian von Neumann entropy of a system is
expressed in terms of the Gaussian (statistical) correlators of
the degrees of freedom of the system.
These correlators are derived from the density matrix
and they are commonly expressed in terms of those fields
that diagonalise the free Hamiltonian. The reason is that
the time evolution of off-diagonal correlators is zero
for non-interacting fields,
which leads to a simple form of the density matrix. Simple means
here that the density matrix can be written as a direct product
of the density matrices for single degrees of freedom.
For interacting theories such as the examples mentioned above,
the off-diagonal correlators (between the different components of the diagonalised Hamiltonian)
are in general nonzero and the density matrix has a more complicated form.
We will discuss this more thoroughly in the coming sections.\\
In our trivial example, the quantum mechanical case for free fermions,
the Hamiltonian is diagonal because there is only one degree of freedom.
Thus the {\it Ansatz}~\eqref{HO:rho:Ansatz} for the
Gaussian density matrix is simple
(and already the most general one),
and so is the expression for the entropy in terms
of the statistical correlator \eqref{HO:entropy:2}.
For a more complicated system such as fermionic fields in $3+1$ dimensions,
which we discuss now,
the system is described by a spinor with four components for every wavenumber.
In order to find the entropy, we first find the fields that diagonalise
the free part of the  Hamiltonian,
then use an {\it Ansatz} for the density matrix in terms of
those fields, and diagonalise it by transforming it into an appropriate
Fock basis.

\subsection{Diagonalisation of the Hamiltonian}
\label{sec: hamiltonian diagonalisation}

As usual, fermions are quantised by employing an anticommutation
relation,
\begin{equation}
 \{\hat\psi_\alpha(\vec x,t),\hat\psi_\beta^\dagger(\vec y,t)\}
        = \delta_{\alpha\beta}\delta^3(\vec x-\vec y\,)
\,,
\label{fermionic quantisation}
\end{equation}
(here $\alpha,\beta\in\{1,2,3,4\}$ are spinor indices, which
are in other equations suppressed)
and thus -- just as in the case of the fermionic quantum mechanics
discussed in
section~\ref{Fermionic quantum mechanics as a 1+0 dimensional field theory} --
 they are Grassmannian operators.

 Upon varying the action \eqref{QFT:action} with respect to $\bar\psi(x)$ and $\psi(x)$,
one gets the following operator equations for $\hat \psi(x)$
and $\hat{\bar\psi}(x)$
,
\begin{equation}
 \imath \gamma^\mu\partial_\mu\hat\psi(x)-m\hat\psi(x) = \hat{j}_\psi(x)
\,,\qquad
 -\imath \partial_\mu\hat{\bar\psi}(x)\gamma^\mu - m\hat{\bar\psi}(x)
               = \hat{\bar j}_\psi(x)
\,,
\label{QFT:operator eom}
\end{equation}
related by hermitian conjugation for real $m$.
The simplest nontrivial case is when the current is generated by
a mixing mass term. In this case
$\hat{j}_\psi(x)=\sum_{j=1}^N m_{0j}\hat\psi_j(x)$ and
$\hat\psi_0(x)\equiv \hat\psi(x)$, $m_{00}\equiv m$, {\it cf.}
Eqs.~(\ref{Yukawa}--\ref{Yukawa:gaussian}).

Here we shall study only time dependent problems,
and we shall work in a spatial cube of volume V,
such that it is convenient to transform these equations into
a spatial momentum space, defined by,
\begin{equation}
 \hat\psi(x) = \frac{1}{\sqrt{V}}\sum_{\vec k} \hat\psi(\vec k,t)
                        {\rm e}^{\imath \vec k\cdot\vec x}
\,;\qquad
 \hat\psi(\vec k,t) = \frac{1}{\sqrt{V}}\int d^3 x  \hat\psi(x)
                        {\rm e}^{-\imath \vec k\cdot\vec x}
\,,
\label{QFT:fourier transforms}
\end{equation}
where $\vec k=(2\pi/L)\vec n$, $L=V^{1/3}$ is the linear size of
the cube $V$,  $\vec n=(n_1,n_2,n_3)$,
$n_i\in {\mathbf Z}$ and ${\mathbf Z}$ is the set of integers.
With these definitions we then get Eqs.~(\ref{fermionic quantisation})
and~(\ref{QFT:operator eom}) in momentum space,
\begin{equation}
 \{\hat\psi_\alpha(\vec k,t),\hat\psi_\beta^\dagger(\vec k^\prime,t)\}
        = \delta_{\alpha,\beta}\delta_ {\vec k,\vec k^\prime}
\,,
\label{fermionic quantisation:2}
\end{equation}
and
\begin{equation}
(\imath \gamma^0\partial_t- \vec\gamma\cdot\vec k-m)\hat\psi(\vec k,t)
   = \hat{j}_\psi(\vec k,t)
\,,\qquad
 -\imath \partial_t\hat{\bar\psi}(\vec k,t)\gamma^0
  +\hat{\bar\psi}(\vec k,t)(\vec k\cdot\vec\gamma-m)
               = \hat{\bar j}_\psi(\vec k,t)
\,,
\label{QFT:operator eom:2}
\end{equation}
where
\begin{equation}
 \hat j_\psi(\vec k,t) = \frac{1}{\sqrt{V}}\int d^3 x  \hat j_\psi(x)
                        {\rm e}^{-\imath \vec k\cdot\vec x}
\,\;\qquad
 \hat{\bar j}_\psi(\vec k,t) = \frac{1}{\sqrt{V}}\int d^3 x
                                \hat{\bar j}_\psi(x)
                                   {\rm e}^{\imath \vec k\cdot\vec x}
\,.
\label{QFT:fourier transforms:j}
\end{equation}
Because the problem at hand is linear, there is no momentum mixing.
Since we are interested in time evolution, we can work in the helicity
eigenbasis, in which
\begin{equation}
\hat\psi(\vec k,t) = \sum_{h=\pm} \hat\psi_h(\vec k,t)\otimes \xi_{h}(\vec k)
\,,\qquad
\hat j_\psi(\vec k,t) = \sum_{h=\pm} \hat j_h(\vec k,t)\otimes \xi_{h}(\vec k)
\,,
\label{helicity decomposition}
\end{equation}
where $\xi_{h}(\vec k)$ are the two-component helicity eigenspinors,
satisfying
\begin{equation}
\hat h \xi_h \equiv \hat{\vec k}\cdot\vec \sigma \xi_h = h \xi_h
\,,
\label{helicity eigenspinors}
\end{equation}
where $\hat h $ is the helicity operator in the two-by-two (Bloch)
representation of Clifford algebra, which can be defined in terms of
the helicity operator $\hat H$ as follows,
$\hat H =\hat{\vec k}\cdot\vec \Sigma = {\rm diag}(\hat h,\hat h)
      =I_2\otimes \hat h$,
$\vec \Sigma = \gamma^0\vec\gamma\gamma^5$.
By making use of the Bloch decomposition of the Clifford algebra
in the Weyl/chiral representation,
\begin{equation}
\gamma^0\rightarrow \rho^1\otimes I
\,,\quad
\gamma^i\rightarrow \imath\rho^2\otimes \sigma^i
\,,\quad
\gamma^5 = \imath \gamma^0\gamma^1\gamma^2\gamma^3
               \rightarrow -\rho^3\otimes I
\,,
\label{Bloch decomposition}
\end{equation}
where $\sigma^{i},\rho^{i}~~(i=1,2,3)$ are the Pauli matrices, equation~\eqref{QFT:operator eom:2}
(when multiplied by $\gamma^0$)
can be rewritten as
\begin{equation}
 (\imath \partial_t+ h k\rho^3 -m\rho^1)\hat\psi_h(\vec k,t)
               = \rho^1\hat{j}_h(\vec k,t)
\,,
\label{QFT:operator eom:3}
\end{equation}
where $\hat\psi_h(\vec k,t)$ is the two spinor whose
components describe the two chiralities $\hat L_h$ and $\hat R_h$
and $k=\|\vec{k}\|$.
Here we shall consider the simpler case when the mass matrix is time
independent.~\footnote{When the mass matrix is time dependent,
$m=m(t)$ and $ \omega=\omega(t)$,
which can occur {\it e.g.} during a phase transition in the early Universe,
then a unitary matrix
is needed to diagonalise the $2\times 2$ problem~\eqref{QFT:operator eom:3},
where now $\theta=\theta(t)$ and $\phi=\phi(t)$.
}
In this case a further (orthogonal) rotation,
\begin{equation}
  R = c_\phi -\imath \rho^2 s_\phi
\,,\quad
     R^T = c_\phi  + \imath \rho^2 s_\phi
\,,\quad
       R\cdot R^T = I = R^T\cdot R
\,,\qquad
     \tan(2\phi)=\frac{m}{hk}
\,,\quad
     \sin(2\phi)=\frac{m}{\omega}
\,,\quad
     \cos(2\phi)=\frac{hk}{\omega}
\,,
\label{QFT:rotaion angles}
\end{equation}
diagonalises equation~(\ref{QFT:operator eom:3}),
where $\omega=\sqrt{k^2+m^2}$,
$c_\phi\equiv \cos(\phi)$ and $s_\phi\equiv \sin(\phi)$.
The resulting (diagonalised) equation~\eqref{QFT:operator eom:3}
is of the form,
\begin{equation}
 (\imath \partial_t+ \omega\rho^3)\hat\Psi_h(\vec k,t)
               = (c_{2\phi}\rho^1-s_{2\phi}\rho^3)\hat{J}_h(\vec k,t)
\,,
\label{QFT:operator eom:diagonal}
\end{equation}
where
\begin{equation}
\hat\Psi_h=R\hat\psi_h=
\left(
\begin{array}{c}
\hat\psi_{h1}\\
\hat\psi_{h2}
\end{array}
\right),
\qquad
\qquad
\hat{J}_h=R\hat j_h=
\left(
\begin{array}{c}
\hat{j}_{h1}\\
\hat{j}_{h2}
\end{array}
\right)
\label{QFT:posnegfreqspinor}
\,,
\end{equation}
and we made use of,
\begin{equation*}
 R\rho^3 R^T = c_{2\phi}\rho^3+s_{2\phi}\rho^1
\,,\qquad
 R\rho^1 R^T = c_{2\phi}\rho^1-s_{2\phi}\rho^3
\,.
\end{equation*}
When Eq.~\eqref{QFT:operator eom:diagonal}
is rewritten in components, we get that
the positive and negative frequency modes (particles and antiparticles)
obey
\begin{align}
    (\imath \partial_t + \omega)\hat\psi_{h1}
             &= -\frac{m}{\omega}\hat j_{h1} +  \frac{hk}{\omega}\hat j_{h2}
\nonumber\\
    (\imath \partial_t - \omega)\hat\psi_{h2}
              &=
    \frac{hk}{\omega}\hat j_{h1}   +   \frac{m}{\omega}\hat j_{h2}
\,.
\label{QFT:operator eom:diagonal2}
\end{align}
In the absence of currents, the problem is reduced to the diagonal one,
and there is no mixing between different states.
We can define a Fock basis $|n^{H}_{h\pm}(\vec k)\rangle$,
 which -- in the absence of interactions --
diagonalises the Hamiltonian. However, in general this procedure does not
diagonalise the density matrix. Only when the source currents vanish and
there is no initial entanglement
between $\hat\psi_{h1}$ and $\hat{\psi}_{h2}$ states
the Fock basis simultaneously diagonalises the
Hamiltonian and density operator. In that case we can define the density operator
 as a direct product
({\it cf.} Eqs.~\eqref{average particle number} and~\eqref{HO:rho:solution:2}) of
the density operators for the different fermionic components.
In general, however, fermionic interactions
(modeled by the currents $\hat j_{h1,2}$) generate
mixing between the $\hat\psi_{h1,2}$ and $\hat j_{h1,2}$ fields, as can be seen
from Eq. \eqref{QFT:operator eom:diagonal2}. Therefore, this mixing should
also be included in the Gaussian density matrix for a fermionic field.

\subsection{Density operator and fermionic entropy}
\label{sec: qft density operator entropy}

Following the previous discussion, a more general {\it Ansatz}
for the Gaussian density operator for an interacting quantum field is
\begin{equation}
\hat\rho(t)=\frac{1}{Z}\exp\Biggl(-\sum_{\vec{k},h}\hat\Psi^{\dagger}_h(\vec{k},t)\Lambda_{h}(k,t)\hat\Psi_h(\vec{k},t) \Biggr)
\,,
\label{QFT: Ansatz density operator}
\end{equation}
where $\hat{\Psi}$ is defined in \eqref{QFT:posnegfreqspinor},
$Z$ is the normalisation constant determined by $\rm{Tr}[\hat\rho]=1$ and
\begin{equation}
\Lambda_{h}(k,t)=
\left(
\begin{array}{cc}
\lambda_{h,11} & \lambda_{h,12}\\
\lambda^{\ast}_{h,12} & \lambda_{h,22}
\end{array}
\right),\qquad \lambda_{h,ij}\equiv \lambda_{h,ij}(k,t)
\,.
\end{equation}
From now on we suppress the momentum labels and time dependence
for the matrix elements $\lambda_{h,ij}$.
Note that the parameters only depend on $k=\|\vec{k}\|$ due to
the assumed spatial homogeneity and isotropy of the state.
In our {\it Ansatz}~\eqref{QFT: Ansatz density operator}
the positive and negative helicity states, as well as the different momentum
states, do not mix,
but the $\hat\psi_{h1,2}$ and $\hat j_{h1,2}$ states do mix through the parameter $\lambda_{12}$.
One could consider adding linear terms in $\hat\Psi,\hat\Psi^{\dagger}$
in \eqref{QFT: Ansatz density operator} that would correspond to a nonzero value of
$\langle \hat{\Psi} \rangle, \langle \hat{\Psi}^{\dagger} \rangle$.
However, if the expectation value is initially zero, it will remain
zero when the system evolves and Eq. \eqref{QFT: Ansatz density operator}
is the most general {\it Ansatz} for $\hat{\rho}$.
The density operator can be diagonalised by a unitary transformation $U$
\begin{equation}
U=\left(
\begin{array}{cc}
\cos\theta & -{\rm e}^{\imath \phi}\sin\theta\\
{\rm e}^{-\imath \phi}\sin\theta & \cos\theta
\end{array}
\right)
\,,
\end{equation}
with $\theta=\theta(t)$ and
\begin{equation}
\cos 2\theta =\frac{\lambda_{11}-\lambda_{22}}{\sqrt{(\lambda_{11}-\lambda_{22})^2+4|\lambda_{12}|^2}},
\qquad\qquad
\sin 2\theta =\frac{-2|\lambda_{12}|}{\sqrt{(\lambda_{11}-\lambda_{22})^2+4|\lambda_{12}|^2}},
\qquad\qquad
{\rm e}^{\imath \phi} =\frac{\lambda_{12}}{|\lambda_{12}|}
\,,
\end{equation}
such that the diagonalised density operator becomes
\begin{equation}
\hat\rho(t)=\frac{1}{Z}
\exp\left(-\sum_{\vec{k},h}\hat{\Psi}^{d\dagger}_h(\vec{k})\Lambda^d_{h}(k,t)\hat{\Psi}^d_h(\vec{k}) \right)
\,,
\label{QFT: diagonal density operator}
\end{equation}
where
\begin{equation}
\hat{\Psi}^d_h(\vec{k})=U\hat\Psi_h(\vec{k},t)=
\left(
\begin{array}{c}
\hat{\psi}_{h+}(\vec{k},t)\\
\hat{\psi}_{h-}(\vec{k},t)
\end{array}
\right),
\qquad
\qquad
\Lambda^d_{h}(k,t)=U\Lambda_{h}(k,t)U^{\dagger}=
\left(
\begin{array}{cc}
\lambda_{h+} & 0\\
0 & \lambda_{h-}
\end{array}
\right),\qquad \lambda_{h\pm}\equiv \lambda_{h\pm}(k,t)
\,.
\end{equation}
Here, we dropped the time dependence in $\hat{\Psi}^{d\dagger}_h(\vec{k})$,
since these are operators in the Schr\"odinger picture.
The eigenvalues are
\begin{align}
\nonumber \lambda_{\pm}&=\frac12(\lambda_{h,11}+\lambda_{h,22})
\pm \frac12\sqrt{(\lambda_{h,11}-\lambda_{h,22})^2+4|\lambda_{h,12}|^2}\\
&= \frac12\rm{Tr}[\Lambda_h]\pm\frac12 \sqrt{(\rm{Tr}[\Lambda_h)^2-4\rm{Det}[\Lambda_h]}
\label{QFT: eigenvalues}
\,.
\end{align}
In the second line the eigenvalues $\lambda_{h\pm}$ are expressed
in terms of the Gaussian invariants of the exponent of the density matrix,
that is, in terms of the trace and determinant of $\Lambda_h=\Lambda_h(k,t)$.
As in the quantum mechanical case, we can identify
the statistical particle number
$\hat N_{h\pm}(\vec k) = \hat{\psi}^\dagger_{h\pm}(\vec k)\hat{\psi}_{h\pm}(\vec k)$
and introduce a Fock basis $|n_{h\pm}(\vec k)\rangle$
(not to be confused with the Fock basis $|n^H_{h\pm}(\vec k)\rangle$
that diagonalises the Hamiltonian, discussed above). Of course
\begin{equation}
\hat N_{h\pm}(\vec k) |n_{h\pm}(\vec k)\rangle
   = n_{h\pm}(k)|n_{h\pm}(\vec k))\rangle
\,,\qquad\qquad
\hat N_{h\pm}(\vec k) = \hat{\psi}^\dagger_{h\pm}(\vec k)\hat{\psi}_{h\pm}(\vec k)
\,.
\label{QFT:stat particle number}
\end{equation}
The trace of the density operator can now be taken easily,
and by demanding $\rm{Tr}[\hat\rho]=1$ the normalisation is
\begin{equation}
Z=\prod_{\vec k, h, \pm} \Bigl(1+{\rm e}^{-\lambda_{h\pm}(k,t)}\Bigr)
\label{QFT: normalisation density operator}
\,.
\end{equation}
Thus the density operator can be written as a direct product,
\begin{equation}
 \hat \rho(t) = \prod_{\vec k,h,\pm} \hat\rho_{h\pm}(\vec k,t)
\,,\qquad
\hat\rho_{h\pm}(\vec k,t) = (1-\bar n_{h\pm}( k,t))
         +(2\bar n_{h\pm}( k,t)-1)\hat N_{h\pm}(\vec k)
\,,
\label{QFT:Fock basis}
\end{equation}
where the average particle number is
\begin{equation}
\bar n_{h\pm}( k,t)=\langle \hat N_{h\pm}(\vec k) \rangle
=\rm{Tr}[\hat{\rho}(t)\hat N_{h\pm}(\vec k)]=[1+\exp(\lambda_{h\pm}(k,t))]^{-1}
\,.
\end{equation}
The entropy is then
({\it cf.} Eq.~(\ref{HO:entropy})),
\begin{equation}
 S = \sum_{\vec kh\pm} s_{h\pm}(k,t)
\,,\qquad
 s_{h\pm}(k,t)
 = -\left(1-\bar n_{h\pm}( k,t)\right)
          \ln\left(1-\bar n_{h\pm}( k,t)\right)
                -\bar n_{h\pm}( k,t)\ln\left(\bar n_{h\pm}( k,t)\right)
\,.
\label{QFT:entropy}
\end{equation}
The main difference between
the quantum fermionic oscillator studied in section~\ref{sec:Fermionic dynamics}
and the (free) quantum field theoretic oscillator
is that for each fermionic mode there are four distinct states (particles),
and hence four contributions to the entropy for each mode $\vec k$:
two from the two helicity states and two from the particle/antiparticle states.
Furthermore,  Eq.~\eqref{QFT:operator eom:diagonal2} implies that
both environmental particles and antiparticles will in general contribute to
the entropy of the system (anti-)particle.
In the non-relativistic limit when $k\rightarrow 0$ particle-antiparticle mixing is absent.
This is a property of the particular form of the interaction \eqref{QFT:action},
which for the purpose of this paper we take to be an operator valued
scalar fermionic current density, which can be generated, for example,
by a mass mixing term.  Other possible interaction Lagrangians that
occur in nature include: pseudo-scalar, vector and pseudo-vector fermionic
currents. For simplicity we shall only consider here the scalar fermionic
current.\\
Just like the quantum mechanical case the average particle numbers
$\bar n_{h\pm}(k,t)$ (and the entropy) can be expressed in terms of statistical correlators.
Using Eqs. \eqref{QFT: Ansatz density operator}, \eqref{QFT: eigenvalues}
and \eqref{QFT: normalisation density operator} one finds
\begin{align}
\nonumber \langle \hat\psi^\dagger_{h1}\hat\psi_{h1}\rangle &
=-\frac{\partial}{\partial \lambda_{h,11}}\ln Z = \frac12 (\bar n_{h+}+\bar n_{h-})
+\frac12 (\bar n_{h+}-\bar n_{h-})
\frac{\lambda_{h,11}-\lambda_{h,22}}{\sqrt{(\lambda_{h,11}-\lambda_{h,22})^2+4|\lambda_{h,12}|^2}}\\
\nonumber \langle \hat\psi^\dagger_{h2}\hat\psi_{h2}\rangle &
=-\frac{\partial}{\partial \lambda_{h,22}}\ln Z = \frac12 (\bar n_{h+}+\bar n_{h-})
-\frac12 (\bar n_{h+}-\bar n_{h-})
\frac{\lambda_{h,11}-\lambda_{h,22}}{\sqrt{(\lambda_{h,11}-\lambda_{h,22})^2+4|\lambda_{h,12}|^2}}\\
\nonumber \langle \hat\psi^\dagger_{h1}\hat\psi_{h2}\rangle &
=-\frac{\partial}{\partial \lambda_{h,12}}\ln Z =
\frac12 (\bar n_{h+}-\bar n_{h-})
\frac{2\lambda_{h,12}^{\ast}}{\sqrt{(\lambda_{h,11}-\lambda_{h,22})^2+4|\lambda_{h,12}|^2}}\\
\langle \hat\psi^\dagger_{h2}\hat\psi_{h1}\rangle &
=-\frac{\partial}{\partial \lambda_{h,12}^{\ast}}\ln Z =
\frac12 (\bar n_{h+}-\bar n_{h-})
\frac{2\lambda_{h,12}}{\sqrt{(\lambda_{h,11}-\lambda_{h,22})^2+4|\lambda_{h,12}|^2}}
\,.
\end{align}
Here $\hat\psi_{hi}=\hat\psi_{hi}(k,t)$, $\bar{n}_{h\pm}=\bar{n}_{h\pm}(k,t)$
and $\lambda_{ij}=\lambda_{h,ij}(k,t)$. We can easily relate the correlators
above to the equal time statistical correlators for the $\hat\psi_{hi}$ fields
\begin{align}
\langle \hat\psi^\dagger_{hi}\hat\psi_{hj}\rangle & =
\frac12\Big\langle\left\{\hat\psi^\dagger_{hi},\hat\psi_{hj}\right\}\Big\rangle
-
\frac12\Big\langle\left[\hat\psi_{hj},\hat\psi^{\dagger}_{hi}\right]\Big\rangle
=\frac12\delta_{ij} -F_{h,ji}
\,,
\end{align}
where $F_{h,ij}=F_{h,ij}(k;t;t)$.
The average particle number expressed in terms of the statistical correlators is then
\begin{equation}
\bar{n}_{h\pm}=\frac12 -\frac12 (F_{h,11}+F_{h,22})\pm
{\rm sgn}(F_{h,22}-F_{h,11})\frac12\sqrt{(F_{h,22}-F_{h,11})^2+4F_{h,12}F_{h,21}}
\label{QFT: particle number statistical correlators}
\,.
\end{equation}
The correctness of this expression can be checked when going to the single quantum mechanical fermion case,
thus only keeping for example the $\hat\psi_{h1}$ fields and setting $\hat\psi_{h2}$ fields to zero.
For specific $h,\vec{k}$ there
is only one remaining particle number, and it agrees with Eq. \eqref{relation particle number stat corr}.
Moreover, in the absence of interactions and with zero initial mixing,
the {\it Ansatz} for the density
matrix~\eqref{QFT: Ansatz density operator} becomes diagonal,
\textit{i.e.} $\lambda_{12}=0$. As we stated earlier in the introduction to this
section,
the density matrix then becomes a direct product of different single fermion
density matrices.
Indeed, the non-interacting case gives $\bar{n}_{h+}=\frac12- F_{h,11}$,
$\bar{n}_{h-}=\frac12- F_{h,22}$, which agrees with the average particle
number for a single fermionic oscillator \eqref{relation particle number stat corr}.
Thus the entropy for a non-interacting fermionic field is simply given by the
sum of the entropies of the components of the diagonalised Hamiltonian, what
was to be expected.
\\
In principle we could have also made an {\it Ansatz}
for the density operator \eqref{QFT: Ansatz density operator}
in terms of rotated fermion fields, for example in terms
of left- and right-handed fields. Of course, the resulting entropy
should not depend on the basis in which the {\it Ansatz} is made,
but one basis may be more convenient than the other.
In order to clarify this, we can define
(just as in the single fermion case) Gaussian invariants of the correlators
\begin{equation}
\Delta_{h\pm}=1-2\bar{n}_{h\pm}=\tanh\Bigl(\frac{\lambda_{h\pm}}{2}\Bigr)
\label{QFT: Gaussian invariant Delta}
\,.
\end{equation}
The fact that the $\Delta_{h\pm}$ are Gaussian invariants of the correlators
becomes more clear when introducing a $2 \times 2$ matrix of statistical correlators
\begin{equation}
{\rm F}_h =
\left(
\begin{array}{cc}
F_{h,11} & F_{h,12}\\
F_{h,21} & F_{h,22}
\end{array}
\right)
\,,
\end{equation}
such that
\begin{equation}
\Delta_{h\pm}={\rm{Tr}}[{\rm{F}}_h]\mp
\sqrt{({\rm{Tr}}[{\rm{F}}_h])^2-4{\rm{Det}}[{\rm{F}}_h]}
\label{QFT: Gaussian invariant correlators}
\,.
\end{equation}
Both the trace and the determinant are invariant under a change of basis,
thus also the expressions for the Gaussian invariant are indeed invariant,
as are the particle number \eqref{QFT: particle number statistical correlators}
and the entropy \eqref{QFT:entropy}. Moreover, because with \eqref{QFT: eigenvalues}
the eigenvalues
$\lambda_{\pm}$ can be expressed in terms of Gaussian invariants of the
density operator,
Eq. \eqref{QFT: Gaussian invariant Delta} presents the
relation between the Gaussian invariants of the correlators and those of
of the density matrix.\\
As a final comment, note that the entropy \eqref{QFT:entropy} only gives
a limited amount of information. The complete density operator
\eqref{QFT: Ansatz density operator} contains more
information and we can separate it into two parts. The "mostly classical"
information is stored in the spectrum, which we define as
\begin{equation}
{\rm Spec}[\hat{\rho}]\equiv \{\lambda_i\}
\label{QFT: spectrum rho}
\,,
\end{equation}
where $\lambda_i$ ($i=1,..,N$) are the eigenvalues of $-\ln(Z\hat{\rho})$
and can be read off from the diagonalised form of the density operator.
To be more precise, for an $N$-state system the components of the spectrum are
\begin{equation*}
\lambda_i=\langle 0_1|..\langle 1_i|..\langle 0_N| (-\ln[Z\hat{\rho}])|0_N\rangle..|1_i\rangle..|0_1\rangle
\,,
\end{equation*}
where here $\hat{\rho}$ is assumed to be written in diagonal form and the $|n_i\rangle$
are the Fock states used to diagonalise the density operator. An
example of the spectrum for a two-state system can be read off
from Eq. \eqref{QFT: diagonal density operator}, where the two
eigenvalues $\lambda_{h\pm}$ \eqref{QFT: eigenvalues} contain
the "mostly classical information". They are related to the
averaged particle numbers $\bar n_{h\pm} =({\rm e}^{\lambda_{h\pm}}+1)^{-1}$,
which is what a late time observer entangled with the Fock states of the system
would identify with a thermal distribution of fermionic particles.
On the other hand, the "mostly quantum" information
is stored in the off-diagonal components of $\hat{\rho}$, which describe
mixing (entanglement) between different states in the original (non-diagonal)
basis. In section~\ref{sec: fermion mass mixing} below we show
both $\{\lambda_i\}$ and $\{\bar n_i\}$.

The Fock states~\eqref{QFT:stat particle number}
are a natural candidate for pointer states
\cite{Zurek:1981xq}, which are selected
by the environment, and in which the system becomes classical,
explaining the above term "mostly classical".  Hence, these Fock
states are particularly useful when considering the process of
classicalization of a quantum system, and in fact we use them to define it.
The rate of statistical particle number increase we identify
as the rate of classicalization~\footnote{Of course, the classicalization rate
is observer dependent and different observers will measure different classicalization rates.
For example, the position operator $\langle \Delta \hat{x}^2 \rangle$
will perceive a different (typically larger) rate of classicalization.}.
The rotation matrix that brings the
density operator to a diagonal form has no classical analogue, thereby
justifying  the name "mostly quantum". Of course, even though the
Fock states~\eqref{QFT:stat particle number} do exist at early times,
the system is then not yet classical. One therefore needs a more precise
definition of when the system becomes classical.

Zurek states~\cite{Zurek:2003zz} that pointer states are
stable under the influence of the environment, but provides no
deeper insight into why this is so. We believe that the stability of
pointer states can be explained by entropic considerations.
Namely, in the limit when the number of environmental oscillators $N$ becomes large,
elements of the reduced density matrix in the diagonal Fock basis become
stable (up to small statistical fluctuations) because most of the volume
of the total Hilbert space (of the system + environment)
corresponds to an almost constant average occupation values of the Fock
states~\eqref{QFT:stat particle number}.
This represents a quantum generalisation of the ergodic hypothesis.

As with regards to classicality, there is a notable difference between bosonic and fermionic
systems. While, in the case of bosonic systems, in the high temperature limit,
one can speak of large occupation numbers of certain oscillators, yielding a
well defined classical field theoretic limit, no classical field theoretic
limit exists for fermions (simply because fermionic occupation numbers
must lie in between 0 and 1). However, that does not mean that there is
no classical limit for fermionic systems.
In the case when the fermionic mass $m$ is very big and
the spatial size of the fermionic system is large
({\it e.g.} when there are many available
system states and they are dense in energy),
one speaks of a classical particle limit,
even when the occupation number of each of the states is much less than 1.
In this case, the number of fermions is well-defined (particle number
fluctuations are suppressed)
and the Pauli blocking is not important.
An important example of the classical particle limit is
the classical fermionic thermal case, in which $m c^2 \gg k_{\rm B} T$,
such that fermions get distributed according to
the Maxwell distribution, which is of course classical.
An analogous classical particle limit exists for bosonic systems as well.

\subsection{Generalisation to $N$ degrees of freedom}
\label{sec: Ndegrees of freedom}

Up to now a density operator $\hat{\rho}$ for two degrees of freedom
(at fixed helicity $h$ and momentum $\vec{k}$) was considered.
Explicit diagonalisation led to a formula for the "phase space"
$\Delta$'s and the entropy \eqref{QFT:entropy}. The diagonal elements
were represented by the invariants ${\rm{Tr}}[{\rm{F}}_h]$ and ${\rm{Det}}[{\rm{F}}_h]$
of the statistical Greens function matrix. Here we consider a more general
setting with $N$ degrees of freedom. The (Gaussian) {\it Ansatz} for $\hat{\rho}$
is now
\begin{equation}
\hat{\rho}=\frac{1}{Z}\exp(-\hat{\psi}^{\dagger}_ia_{ij}\hat{\psi}_j)
\,, \qquad\qquad
(i,j=1,..,N)
\,.
\label{NDOF: Ansatz rho}
\end{equation}
Indeed this is equivalent to a $\rho(\bar{\theta}',\theta)$ in the
coherent state representation:
\begin{equation}
\rho(\bar{\theta}',\theta)=\langle \theta'| \hat{\rho} | \theta \rangle
= \frac{1}{Z}\exp(\bar{\theta}^{\prime}_i M_{ij} \theta_j)
\,,
\label{NDOF: rhocoherentrepresentation}
\end{equation}
with
\begin{align}
\nonumber M_{ij}&=\left({\rm e}^{-a}\right)_{ij}\\
Z&={\rm{Det}}[\mathbb{I}+{\rm e}^{-a}]
\,,
\end{align}
as one can easily see by diagonalising the hermitian (real, symmetric)
matrices $a,M$ simultaneously (as in the $2\times2$ case considered before).
The 2-correlators $\langle \hat{\psi}^{\dagger}_k \hat{\psi}_l\rangle$
are related to the statistical "matrix" ${\rm{F}}_{kl}$ and to the $a_{ik}$
introduced above:
\begin{equation}
\bar{n}_{kl}\equiv\langle \hat{\psi}^{\dagger}_k \hat{\psi}_l\rangle =
\frac12\delta_{kl} -{\rm{F}}_{lk}(t;t)=\frac12 (\mathbb{I}-\Delta)_{kl}
\,,
\end{equation}
which is a suitable generalisation to many degrees of freedom
of the one degree of freedom result~\eqref{relation particle number stat corr}.
$\langle \hat{\psi}^{\dagger}_k \hat{\psi}_l\rangle$
can be also obtained by differentiating ${\rm{Tr}}[\hat{\rho}]$ of
 Eq. \eqref{NDOF: Ansatz rho} with respect to $-a_{kl}$:
\begin{equation}
-\frac{\partial}{\partial a_{kl}} {\rm{Tr}}[\hat{\rho}] = 0 =
{\rm{Tr}}\left[\hat{\psi}^{\dagger}_k \hat{\psi}_l\frac{\exp(-\hat{\psi}^{\dagger}_ia_{ij}\hat{\psi}_j)}{Z}\right]
-{\rm{Tr}}\left[\exp(-\hat{\psi}^{\dagger}_ia_{ij}\hat{\psi}_j)\right]\frac{\partial}{\partial a_{kl}}
\left(\frac{1}{Z}\right)
\,,
\end{equation}
where the first term is just $\langle \hat{\psi}^{\dagger}_k \hat{\psi}_l\rangle$
and the second term is evaluated as
\begin{align}
\nonumber -Z\frac{\partial}{\partial a_{kl}}\left(\frac{1}{Z}\right) & =
\frac{1}{Z}\frac{\partial}{\partial a_{kl}}{\rm{Det}}[\mathbb{I}+{\rm e}^{-a}]\\
\nonumber & = \frac{\partial}{\partial a_{kl}}{\rm Tr}\ln\left(\mathbb{I}+{\rm e}^{-a}\right)\\
&= -\left(\frac{{\rm e}^{-a}}{\mathbb{I}+{\rm e}^{-a}}\right)_{lk}
= - \left(\frac{\mathbb{I}}{\mathbb{I}+{\rm e}^{a}}\right)_{lk}
\,.
\end{align}
We have used here the identity ${\rm Det}[A]=\exp\{{\rm tr} [\ln(A)]\}$,
and that $(\partial/\partial a_{ij}){\rm Tr}{f(a)}=\left(f'(a)\right)_{ji}$,
where $f$ is some function of the matrix $a_{ij}$.
We thus obtain the generalisation of Eqs.
\eqref{relation particle number stat corr},
\begin{equation}
\bar{n}_{kl}\equiv\frac12 (\mathbb{I}-\Delta)_{kl}=\left[(\mathbb{I}+{\rm e}^{a})^{-1}\right]_{lk}
\,.
\label{NDOF: generalisationDelta}
\end{equation}
Diagonalising $a, \Delta, {\rm F}$ with the same rotation we can use
a sum of terms of type Eq. \eqref{HO:entropy:2} for the entropy.
This is a trace and rotating back inside the trace we obtain
\begin{equation}
S=-{\rm Tr}\left[\frac{\mathbb{I}+\Delta}{2}\ln\Bigl(\frac{\mathbb{I}+\Delta}{2}\Bigr)
      +\frac{\mathbb{I}-\Delta}{2}\ln\Bigl(\frac{\mathbb{I}-\Delta}{2}\Bigr)\right]=
      -{\rm Tr}\left[(\mathbb{I}-\bar{n})\ln(\mathbb{I}-\bar{n})
      +\bar{n}\ln\bar{n}\right]
\,.
\label{NDOF: generalisationEntropy}
\end{equation}
For this formula to work the eigenvalues of $\bar{n}$ must lie in the interval
$[0,1]$, which is indeed the case for fermionic systems.
The result~\eqref{NDOF: generalisationEntropy}
we have also obtained using the replica trick for the density operator
$\hat{\rho}$ in the coherent
representation \eqref{NDOF: rhocoherentrepresentation}
without a diagonalisation procedure.
Appendix \ref{Appendix D: Entropy via the replica trick in coherent state basis}
contains the calculational details.
Of course in order to evaluate
\eqref{NDOF: generalisationEntropy}, 
diagonalisation of $\Delta, a$ is again the fastest method to obtain
the entropy. In the next section we discuss the growth of entropy for
Dirac fermions mixing through a mass matrix, where we will also
show the spectrum of $\hat{\rho}$ \eqref{QFT: spectrum rho}
as well as the average particle number.

\subsection{Fermion mass mixing}
\label{sec: fermion mass mixing}

A simple model for interacting fermions is a model of
different fermion species mixing through a mass matrix.
Similar to the quantum mechanical action of
bilinearly coupled fermions \eqref{BCF:bilinearfermionmodel},
the action for fermion mass mixing
\begin{equation}
S[\hat{\psi}_x,\{\hat{\psi}_{q_i}\}]=\int d^4x \left\{ \mathcal{L}_{\text{S}}[\hat{\psi}_x]
+ \mathcal{L}_{\text{E}}[\{\hat{\psi}_{q_i}\}] +
\mathcal{L}_{\text{int}}[\hat{\psi}_x,\{\hat{\psi}_{q_i}\}]\right\}
\,,
\label{QFT:FMM:fermionmassmixingmodel}
\end{equation}
with
\begin{align}
\nonumber \mathcal{L}_{\text{S}}[\hat{\psi}_x]&=
\hat{\psi}_x^{\dagger}(x) (\imath\gamma^{\mu}\partial_{\mu}-m_0)\hat{\psi}_x(x)\\
\nonumber \mathcal{L}_{\text{E}}[\{\hat{\psi}_{q_i}\}]&=
\sum_{i=1}^N\hat{\psi}_{q_i}^{\dagger}(x) (\imath\gamma^{\mu}\partial_{\mu}-m_i)\hat{\psi}_{q_i}(x)\\
\mathcal{L}_{\text{int}}[\hat{\psi}_x,\{\hat{\psi}_{q_i}\}]&=
-\sum_{i=1}^N\left(m_{0i}\hat{\psi}_x^{\dagger}\hat{\psi}_{q_i} +m_{i0}\hat{\psi}_{q_i}^{\dagger}\hat{\psi}_x\right)
\,.
\end{align}
From now on we assume that the mass mixing parameters are real, $m_{0i}=m_{i0}$.
Next, we follow the same steps as in section~\ref{sec: hamiltonian diagonalisation}:
we first transform all fields to momentum space as in \eqref{QFT:fourier transforms},
then go to the helicity eigenbasis using \eqref{helicity decomposition},
and finally rotate the fields as in Eq. \eqref{QFT:posnegfreqspinor}
(with a different rotation matrix $R_i$ for different species).
The resulting action is
\begin{align}
\nonumber S[\hat{\psi}_x,\{\hat{\psi}_{q_i}\}]=\int dt \sum_{\vec{k},h}\Biggl\{&
\hat{\Psi}_{x,h}^{\dagger}(\vec{k},t)\left(i\partial_t+\omega_0\rho^3\right)\hat{\Psi}_{x,h}(\vec{k},t)
+\sum_{i=1}^N\hat{\Psi}_{q_{i},h}^{\dagger}(\vec{k},t)\left(i\partial_t+\omega_i\rho^3\right)\hat{\Psi}_{q_{i},h}(\vec{k},t)\\
&-\sum_{i=1}^N m_{0i} \left[\hat{\Psi}_{x,h}^{\dagger}
R\rho^1 R_i^T\hat{\Psi}_{q_i,h}
+\hat{\Psi}_{q_i,h}^{\dagger}R_i\rho^1 R^T\hat{\Psi}_{x,h}\right]
\Biggr\}
\label{QFT:FMM: diagonalaction}
\,,
\end{align}
where $\omega_0=\sqrt{m_0^2+\|\vec{k}\|^2}$ and
$\omega_i=\sqrt{m_i^2+\|\vec{k}\|^2}$.
$R=R(\theta)$ is the rotation matrix
that diagonalises the free part of the action for $\hat{\psi}_x$,
whereas $R_i=R(\theta_i)$ diagonalises the action for $\hat{\psi}_{q_i}$.
When the masses of different species are the same, $m_0=m_i$, the
rotation matrices will be the same, $R=R_i$. In general
the interaction term is
\begin{equation}
R\rho^1 R_i^T=R_i\rho^1 R^T=\cos(\theta+\theta_i)\rho^1-\sin(\theta+\theta_i)\rho^3
\,,
\end{equation}
with the $\theta,\theta_i$ defined as in Eq. \eqref{QFT:rotaion angles}.
%
The equations of motions follow directly from the action
\eqref{QFT:FMM: diagonalaction}

\begin{align}
\nonumber (\imath\partial_t+\omega_0\rho^3)\hat{\Psi}_{x,h}(\vec{k},t)&
=\sum_{i=1}^N m_{0i}R\rho^1 R_i^T\hat{\Psi}_{q_i,h}(\vec{k},t)\\
\nonumber(\imath\partial_t+\omega_i\rho^3)\hat{\Psi}_{q_i,h}(\vec{k},t)&
=m_{0i}R_i\rho^1 R^T\hat{\Psi}_{x,h}(\vec{k},t)\\
\nonumber(-\imath\partial_t\hat{\Psi}^{\dagger}_{x,h}(\vec{k},t)
+\omega_0 \hat{\Psi}^{\dagger}_{x,h}(\vec{k},t)\rho^3)&
=\sum_{i=1}^N m_{0i}\hat{\Psi}^{\dagger}_{q_i,h}(\vec{k},t)R_i\rho^1 R^T\\
(-\imath\partial_t\hat{\Psi}^{\dagger}_{q_i,h}(\vec{k},t)
+\omega_i\hat{\Psi}^{\dagger}_{q_i,h}(\vec{k},t)\rho^3)&
=m_{0i}\hat{\psi}^{\dagger}_{x,h}(\vec{k},t)R\rho^1 R_i^T
\,.
\label{QFT:FMM:eompsix}
\end{align}
Note that each line consists of two equations for the two components
of the spinors $\hat{\Psi}_{x,h}$ and $\hat{\Psi}_{q_i,h}$.
Differential equations for the statistical equal-time correlators
can be derived from the equations of motions \eqref{QFT:FMM:eompsix}.
It is only necessary to derive the statistical correlators
for fields with the same helicity. Remember that
there is no helicity mixing in the action.
For the system alone, the statistical correlators obey,
\begin{align}
\nonumber \imath\partial_t F_{xx,h11}&
=\sum_{i=1}^N m_{0i}\left[\cos(\theta+\theta_i)\left(F_{q_ix,h21}-F_{xq_i,h12}\right)
-\sin(\theta+\theta_i)\left(F_{q_ix,h11}-F_{xq_i,h11}\right)\right]\\
\nonumber \imath\partial_t F_{xx,h22}&
=\sum_{i=1}^N m_{0i}\left[\cos(\theta+\theta_i)\left(F_{q_ix,h12}-F_{xq_i,h21}\right)
+\sin(\theta+\theta_i)\left(F_{q_ix,h22}-F_{xq_i,h22}\right)\right]\\
(\imath\partial_t+2\omega_0) F_{xx,h12}&
=\sum_{i=1}^N m_{0i}\left[\cos(\theta+\theta_i)\left(F_{q_ix,h22}-F_{xq_i,h11}\right)
-\sin(\theta+\theta_i)\left(F_{q_ix,h12}+F_{xq_i,h12}\right)\right]
\,.
\label{QFT:FMM:eomFxx}
\end{align}
Here we have used a shorthand notation, with
\begin{align}
F_{xx,hmn}&\equiv F_{xx,hmn}(k;t;t)
=\frac12\langle [\hat{\psi}_{x,hm}(\vec{k},t),\hat{\psi}^{\dagger}_{x,hn}(\vec{k},t)]\rangle
\,,
\qquad\qquad
m,n=1,2
\,,
\qquad\qquad
\rm{etc}
\,.
\end{align}
Note that $F^{\star}_{xx,hmn}=F_{xx,hnm}$, $F^{\star}_{xq_i,hmn}=F_{q_ix,hnm}$
and $F^{\star}_{q_iq_j,hmn}=F_{q_jq_i,hnm}$. With these relations the remaining
equation for $F_{xx,h21}$ in \eqref{QFT:FMM:eomFxx}
can be found easily by taking the complex conjugate.
Similarly, the equations of motion for the environmental correlators are:
\begin{eqnarray}
 (\imath\partial_t+(\omega_i-\omega_j)) F_{q_iq_j,h11}
&=& m_{0i}\left[\cos(\theta+\theta_i)F_{xq_j,h21}
-\sin(\theta+\theta_i)F_{xq_j,h11}\right]
\nonumber\\
&&-\,m_{0j}\left[\cos(\theta+\theta_j)F_{q_ix,h12}
   -\sin(\theta+\theta_j)F_{q_ix,h11}\right]
\nonumber\\
 (\imath\partial_t-(\omega_i-\omega_j)) F_{q_iq_j,h22}
&=& m_{0i}\left[\cos(\theta+\theta_i)F_{xq_j,h12}
+\sin(\theta+\theta_i)F_{xq_j,h22}\right]
\nonumber\\
&&-\,m_{0j}\left[\cos(\theta+\theta_j)F_{q_ix,h21}
+\sin(\theta+\theta_j)F_{q_ix,h22}\right]
\nonumber\\
(\imath\partial_t+(\omega_i+\omega_j)) F_{q_iq_j,h12}
&=& m_{0i}\left[\cos(\theta+\theta_i)F_{xq_j,h22}
-\sin(\theta+\theta_i)F_{xq_j,h12}\right]
\nonumber\\
&&-\,m_{0j}\left[\cos(\theta+\theta_j)F_{q_ix,h11}
+\sin(\theta+\theta_j)F_{q_ix,h12}\right]
\,.
\label{QFT:FMM:eomFqq}
\end{eqnarray}
Again, the remaining equation for $F_{q_iq_j,h21}$ can
be obtained by complex conjugation of the third line above.
Finally, the system-environment correlators obey the equations
\begin{eqnarray}
 (\imath\partial_t+(\omega_0-\omega_i)) F_{xq_i,h11}
&=& \sum_{j=1}^N m_{0j}\left[\cos(\theta+\theta_j)F_{q_jq_i,h21}
-\sin(\theta+\theta_j)F_{q_jq_i,h11}\right]
\nonumber\\
&&-\,m_{0i}\left[\cos(\theta+\theta_i)F_{xx,h12}
-\sin(\theta+\theta_i)F_{xx,h11}\right]
\nonumber\\
 (\imath\partial_t-(\omega_0-\omega_i)) F_{xq_i,h22}
&=&  \sum_{j=1}^N m_{0j}\left[\cos(\theta+\theta_j)F_{q_jq_i,h12}
+\sin(\theta+\theta_j)F_{q_jq_i,h22}\right]
\nonumber\\
&&-\,m_{0i}\left[\cos(\theta+\theta_i)F_{xx,h21}
+\sin(\theta+\theta_i)F_{xx,h22}\right]
\nonumber\\
 (\imath\partial_t+(\omega_0+\omega_i)) F_{xq_i,h12}
&=& \sum_{j=1}^N m_{0j}\left[\cos(\theta+\theta_j)F_{q_jq_i,h22}
-\sin(\theta+\theta_j)F_{q_jq_i,h12}\right]
\nonumber\\
&&-\,m_{0i}\left[\cos(\theta+\theta_i)F_{xx,h11}
+\sin(\theta+\theta_i)F_{xx,h12}\right]
\nonumber\\
(\imath\partial_t-(\omega_0+\omega_i)) F_{xq_i,h21}
&=& \sum_{j=1}^N m_{0j}\left[\cos(\theta+\theta_j)F_{q_jq_i,h11}
+\sin(\theta+\theta_j)F_{q_jq_i,h21}\right]
\nonumber\\
&&-\,m_{0i}\left[\cos(\theta+\theta_i)F_{xx,h22}
-\sin(\theta+\theta_i)F_{xx,h21}\right]
\,.
\label{QFT:FMM:eomFxq}
\end{eqnarray}
Taking the complex conjugate of these equations
gives the final equations of motion for
the environment-system correlators. This
results in a closed system of $(N+1)^2\times 2^2\times 2$
equations for the correlators of the components of
$(N+1)$ coupled 2-spinors at different helicities.
These coupled first order differential equations
can be solved (numerically) with initial conditions
corresponding to environmental oscillators in chemical equilibrium:
\begin{align}
\nonumber F_{xx,h11}(k;t_0;t_0)&=F_{xx,h22}(k;t_0;t_0)=\frac{1}{2}\\
\nonumber F_{q_iq_j,h11}(k;t_0;t_0)&=
\delta_{ij}\frac{1}{2}\tanh\left(\frac{\beta(\omega_i-\mu_i)}{2}\right)\\
F_{q_iq_j,h22}(k;t_0;t_0)&=
\delta_{ij}\frac{1}{2}\tanh\left(\frac{\beta(\omega_i+\mu_i)}{2}\right)
\,,
\label{QFT:FMM:initialconditions}
\end{align}
and all others are initially equal to zero. According to
Eqs. \eqref{QFT:FMM:initialconditions} at $t=t_0$
there is no mixing between the different components
of the 2-spinors in the helicity eigenbasis. Remember
that these '1,2' components are the fields that diagonalise
the Hamiltonian in a non-interacting theory; they are the
positive and negative frequency states,
or particles and antiparticles. Note that
the initial state~(\ref{QFT:FMM:initialconditions}) allows
for nonvanishing chemical potentials $\mu_i$
for the environmental fields. The chemical potentials have an opposite sign
for particles and antiparticles.
Moreover, we have assumed that initially there is no mixing between
the system and the environment. The physical picture
is therefore that initially there is no mass-mixing between
the different fermion species, but at $t=t_0$ the coupling
is switched on and the entropy of the system can grow.
That is, we consider only the entropy of the system
\begin{align}
\nonumber S_x(t)&=\sum_{\vec{k}h\pm}s_{xx,h\pm}(k,t)\\
&=\sum_{\vec{k}h\pm}\left[-\frac{1+\Delta_{xx,h\pm}(k,t)}{2}\ln\left(\frac{1+\Delta_{xx,h\pm}(k,t)}{2}\right)
-\frac{1-\Delta_{xx,h\pm}(k,t)}{2}\ln\left(\frac{1-\Delta_{xx,h\pm}(k,t)}{2}\right)\right]
\,,
\label{QFT:FMM:entropysytem}
\end{align}
where $s_{xx,h\pm}(\|\vec k\|,t)$
is the system entropy per fermionic degree of freedom, {\it i.e.} for a state
with quantum numbers $\vec k,h,\pm$.
The Gaussian invariants $\Delta_{xx,h\pm}$ are
those defined in Eq. \eqref{QFT: Gaussian invariant correlators},
with the subscript $xx$ indicating that only the system correlators
are used. Due to the loss of information (assuming environmental
correlations are inaccessible) the system decoheres, leading
to an increase in entropy for the system.\\
\begin{figure}[t!]
    \begin{minipage}[t]{.45\textwidth}
        \begin{center}
\includegraphics[width=\textwidth]{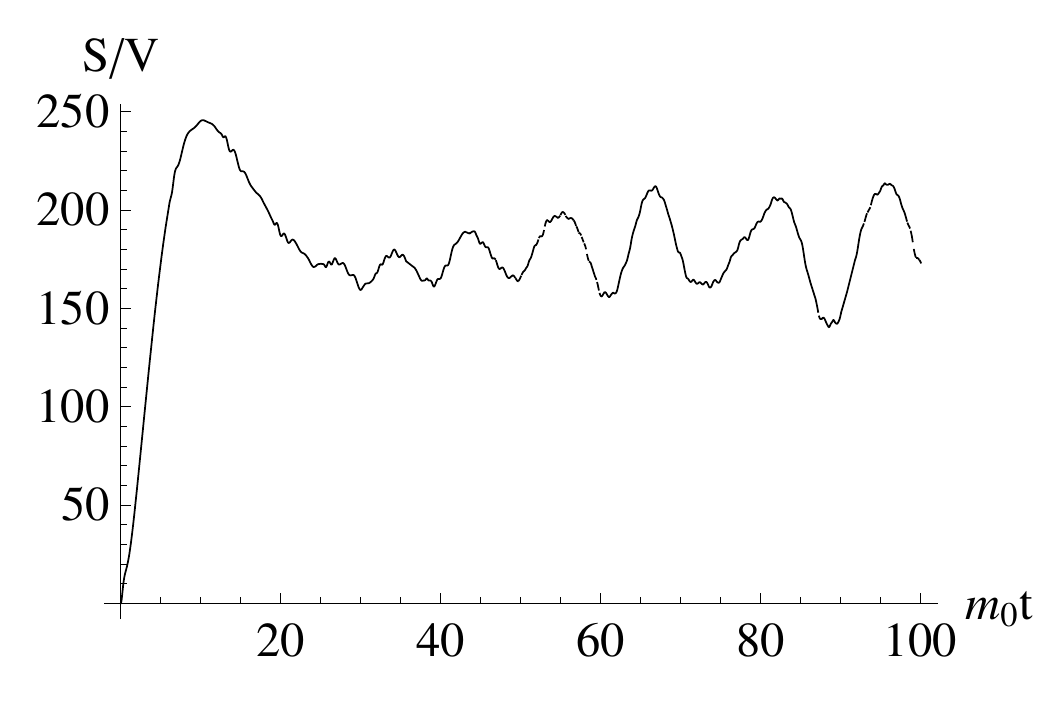}
   {\em \caption{System entropy as a function of $m_0 t$ for $N=1$ environmental field
   and zero chemical potential, $\mu_1=0$.
     The parameters are $m_1=1.1 m_0$, $m_{01}=0.5 m_0$ and $\beta=(m_0)^{-1}$.
     The entropy is expressed per $\left(\frac{m_0}{2\pi}\right)^3$.
      \label{fig:QFT:2coupledfermions1} }}
        \end{center}
    \end{minipage}
\hfill
    \begin{minipage}[t]{.45\textwidth}
        \begin{center}
\includegraphics[width=\textwidth]{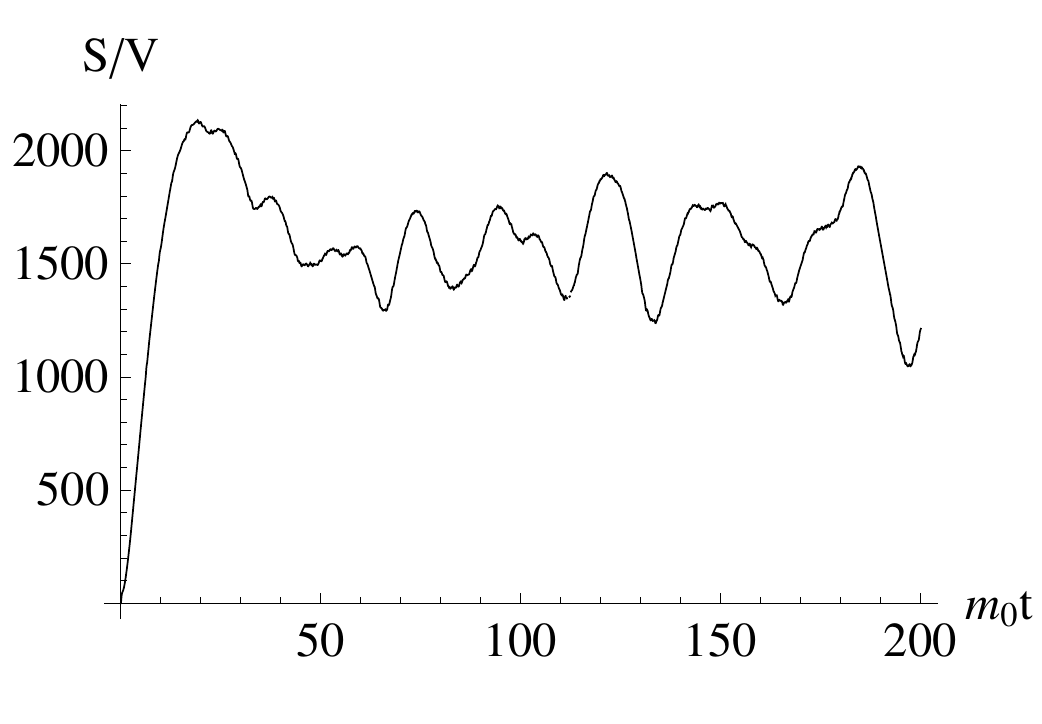}
   {\em \caption{System entropy as a function of $m_0 t$ for $N=1$ environmental field
   and zero chemical potential, $\mu_1=0$.
   The parameters are the same as those in Fig. \ref{fig:QFT:2coupledfermions1},
   but the temperature is twice higher, $\beta=0.5(m_0)^{-1}$.
   The late time entropy is approximately 8 times higher than for $\beta=(m_0)^{-1}$,
   which supports the scaling of late time entropy as $T^3$.
   \label{fig:QFT:2coupledfermions2} }}
        \end{center}
    \end{minipage}
\end{figure}
In the case of $N=1$ environmental fields we have
numerically solved the $16\times 2$ equations for the statistical
correlators of particles and antiparticles of the system and environment
at different helicities. In the numerical procedure a smooth selection of
modes $\vec k$ has been made, separated into spherical bins of size $\Delta k$.
The total system entropy~\eqref{QFT:FMM:entropysytem}
is then calculated as
\begin{equation*}
S_x(t)=\sum_{\vec kh\pm}s_{xx,h\pm}(k,t)
      = V\left(\frac{m_0}{2\pi}\right)^3
  \sum_{k/\Delta k=0}^\infty\sum_{h=\pm}\sum_{\pm}
       4\pi \left(\frac{k}{m_0}\right)^2
          \left(\frac{\Delta k}{m_0}\right) s_{xx,h\pm}(k,t)
\,,
\end{equation*}
where $V$ is the volume of the system and
$s_{xx,h\pm}(k,t)$ is the (average) entropy per degree of freedom
in a spherical bin with a momentum
$\|\vec k\|$ and a width $\Delta k\ll k$. The maximum mode has been
chosen such that $\beta k_{\rm{max}}\gg 1$, since the inclusion
of higher modes does not significantly change the total entropy.
In Figs. \ref{fig:QFT:2coupledfermions1}
and \ref{fig:QFT:2coupledfermions2} the entropy density for the system
(in units of the inverse Compton wavelength cubed,
$\lambda_{\rm C}^{-3}=({m_0}/{(2\pi)})^3$)
has been plotted for zero chemical potential, same mass and mass-mixing
parameters, but different temperatures. In the absence of a chemical potential
the particles and antiparticles evolve completely separately, \textit{i.e.}
$F_{xx,h12}$ and $F_{xx,h21}$ are zero. Because the
initial conditions are identical the statistical correlators
$F_{xx,h11}$ and $F_{xx,h22}$ behave equally, and so do the Gaussian
invariants $\Delta_{xx,h\pm}$ of Eq. \eqref{QFT: Gaussian invariant correlators}.
Thus the total entropy is simply four times the entropy calculated from
a single Gaussian invariant. In Figs. \ref{fig:QFT:2coupledfermions1}
and \ref{fig:QFT:2coupledfermions2} the total system entropy increases
due to interactions with the environment. After some time it fluctuates
around an equilibrium value. This late time entropy should scale
as $\beta^{-3}=(k_{{\rm B}}T)^3$ in the relativistic limit where
$k_{{\rm B}}T/m_0=1/(\beta m_0)\gg 1$. Comparing Figs. \ref{fig:QFT:2coupledfermions1}
and \ref{fig:QFT:2coupledfermions2} this appears to be the case.\\
The fluctuations in late time entropy in Figs.~\ref{fig:QFT:2coupledfermions1}
and \ref{fig:QFT:2coupledfermions2} are rather large. The reason is that the bilinear
coupling is not a true interaction term: each system field mode is only coupled
to $N=1$ environmental mode. Due to the unitary evolution, energy flows back and forth
from the system to the environmental oscillator, resulting in large amplitude oscillations.
As mentioned at the end of
Sec.~\ref{N+1 coupled fermionic oscillators}, in the case of a true interaction each
system mode couples effectively to infinitely many environmental modes,
leading to an efficient thermalization of the system with an expected rate that is
to a good approximation given by the perturbative rate, just as
in the case of bosonic field theory~\cite{Koksma:2011dy}.\\
\begin{figure}[t!]
    \begin{minipage}[t]{.45\textwidth}
        \begin{center}
\includegraphics[width=\textwidth]{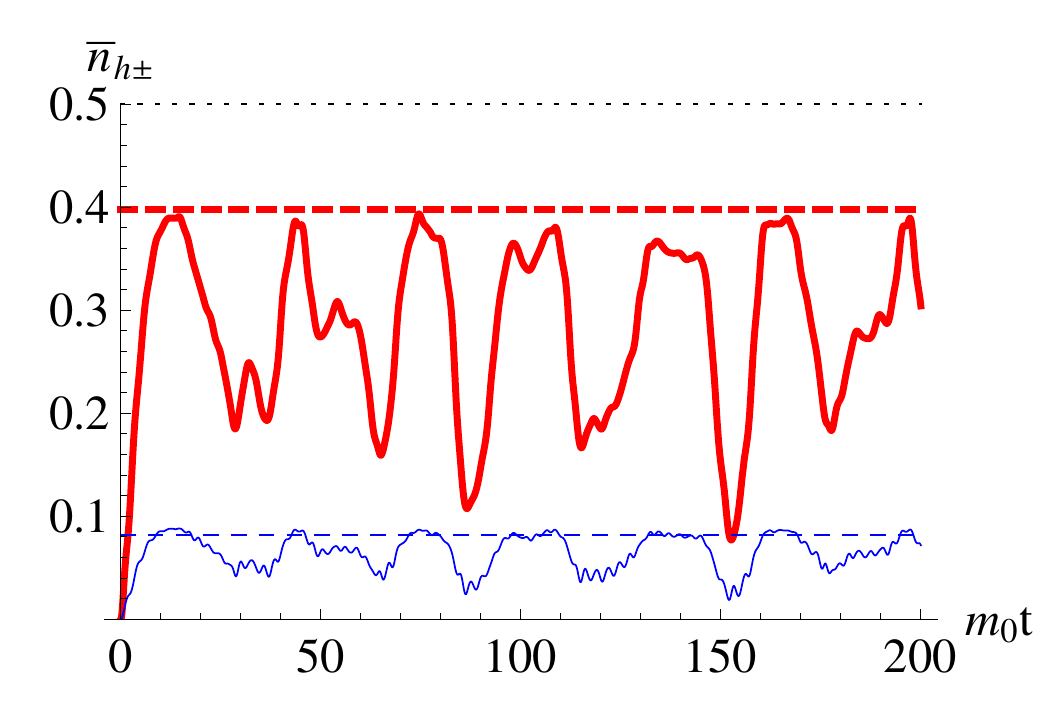}
   {\em \caption{Average (anti)particle number $\bar{n}_{h\pm}$
   for the mode $k=m_0$ as a function of $m_0 t$ for $N=10$ environmental fields.
   The environmental masses are distributed as $m_i=(0.5i-0.25)\times m_0$, or
   $m_1=0.25m_0, m_2=0.75m_0,.., m_{10}=4.75m_0$, and the couplings are all equal
   $m_{0i}=0.2 m_0$. The inverse temperature is $\beta=(m_0)^{-1}$
   and the (equal) chemical potentials are $\mu_i=m_0$.
   Due to the chemical potential the antiparticle number
   density $\bar{n}_{h-}$ (solid blue) is suppressed with respect
   to the particle number density $\bar{n}_{h+}$ (solid red, thick).
   The dashed lines indicate the (anti)particle number densities for
   perfect thermalisation,
   $(\bar{n}_{{\rm th}})_{h\pm}=({\rm e}^{\beta(\omega\mp\mu_1)}+1)^{-1}$.
   Note that there is no distinction in particle number for
   $+$ and $-$ helicity states because helicity mixing is absent.
   The dotted black line is the maximum fermionic particle number
   in the limit when $\beta\rightarrow 0$, $\bar{n}_{h\pm}\rightarrow 0.5$.
      \label{fig:QFT:particlenumber1} }}
        \end{center}
    \end{minipage}
\hfill
    \begin{minipage}[t]{.45\textwidth}
        \begin{center}
\includegraphics[width=\textwidth]{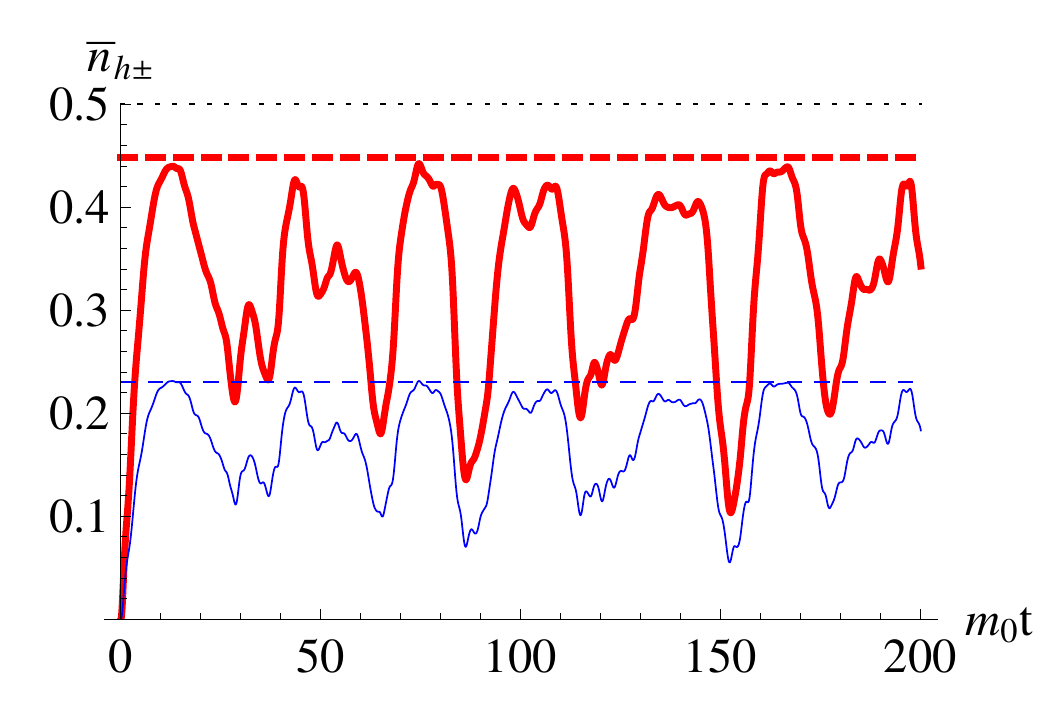}
   {\em \caption{Average (anti)particle number $\bar{n}_{h\pm}$
   for the mode $k=m_0$ as a function of $m_0 t$ for $N=10$ environmental fields.
   The parameters are the same as those in Fig. \ref{fig:QFT:particlenumber1},
   but the temperature is higher, $\beta=0.5(m_0)^{-1}$. Both the particle (solid red, thick)
   and antiparticle (solid blue) number densities are larger than in
   Fig. \ref{fig:QFT:particlenumber1}, but the relative increase of the antiparticle
   number density is bigger.
      \label{fig:QFT:particlenumber2} }}
        \end{center}
    \end{minipage}
\end{figure}
\begin{figure}[t!]
    \begin{minipage}[t]{.45\textwidth}
        \begin{center}
\includegraphics[width=\textwidth]{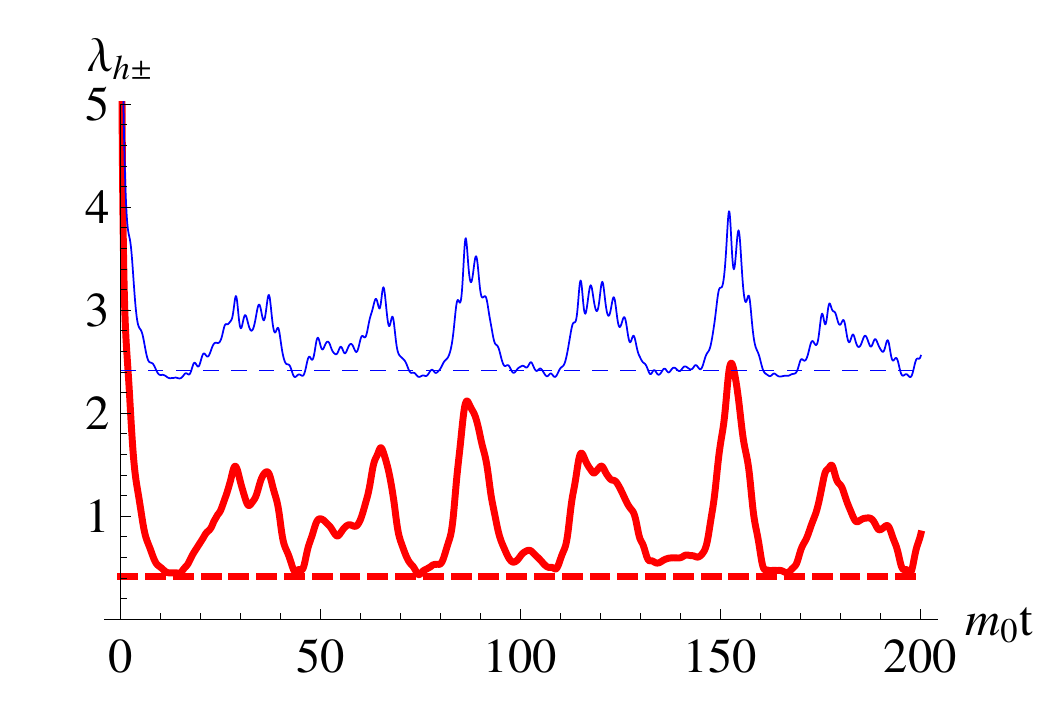}
   {\em \caption{Spectrum of the density operator for the mode $k=m_0$
   as a function of $m_0 t$ for $N=10$ environmental fields.
   The spectrum, defined in Eq. \eqref{QFT: spectrum rho}, are
   the $\lambda_{h\pm}$ of Eq. \eqref{QFT: Gaussian invariant Delta}.
   The parameters are the same as those in Fig. \ref{fig:QFT:particlenumber1}.
   Due to the chemical potential $\lambda_{h-}$ (solid blue) is greater
   than $\lambda_{h+}$ (solid red, thick).
   When the system is completely thermalised the spectrum is
   $\lambda_{h\pm}=\beta(\omega_0\mp\mu_1)$, indicated by the red and blue dashed
   lines for particles and antiparticles, respectively.
   Note that the initial value of $\lambda_{h\pm}$ is infinite
   as the initial particle number is zero.
   \label{fig:QFT:spectrum1} }}
        \end{center}
    \end{minipage}
\hfill
    \begin{minipage}[t]{.45\textwidth}
        \begin{center}
\includegraphics[width=\textwidth]{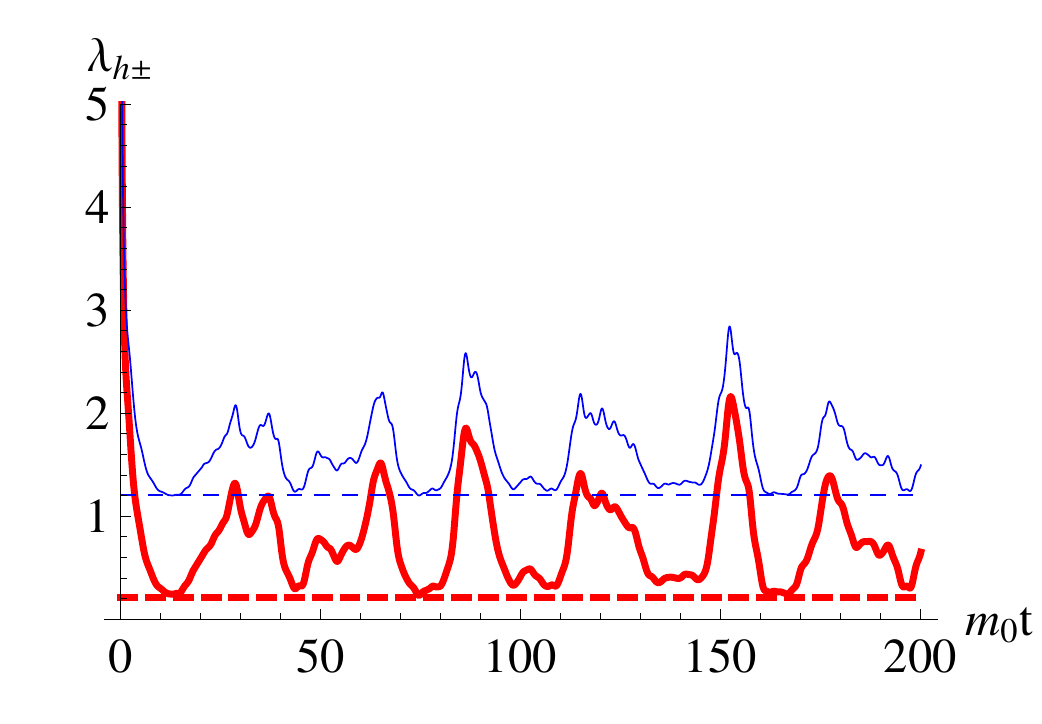}
   {\em \caption{Spectrum of the density operator for the mode $k=m_0$
   as a function of $m_0 t$ for $N=10$ environmental fields.
   The parameters are the same as those in Fig. \ref{fig:QFT:spectrum1},
   but the temperature is higher, $\beta=0.5(m_0)^{-1}$. As temperature
   increases, the difference between $\lambda_{h+}$ and $\lambda_{h-}$ becomes
   smaller.
   \label{fig:QFT:spectrum2} }}
        \end{center}
    \end{minipage}
\end{figure}
Next, a distinction can be made between particles
and antiparticles by introducing a nonzero chemical potential $\mu_i$
for the environmental fermion species $\hat{\psi}_{q_i}$.
In Figs. \ref{fig:QFT:particlenumber1}--\ref{fig:QFT:spectrum2}
the particle/antiparticle number densities
\eqref{QFT: particle number statistical correlators}
and the spectrum of the density operator \eqref{QFT: spectrum rho}
are shown at different values of $\beta$.
We have taken here the case of a system field interacting with $10$
environmental fields, which have masses distributed around the
system mass $m_0$.
In general,
the (anti)particle number oscillates between the initial value $0$
and the value for perfect thermalisation, when the system fermions
have the same temperature as the environmental fermions,
approximately the initial temperature of the environment.
Moreover, Figs. \ref{fig:QFT:particlenumber1}
and \ref{fig:QFT:particlenumber2} clearly show that,
for positive particle environmental chemical potentials, the system antiparticle
number density is suppressed with respect to the particle number density
due to the nonzero chemical potentials. As usual, for higher temperatures
(lower $\beta$) the particle numbers are closer to the maximum fermionic
particle number $\bar{n}_{\rm max}$ for $\beta\rightarrow 0$, but the relative
increase of the antiparticle number with respect to particle number is greater.\\
The spectrum of the density operator \eqref{QFT: spectrum rho} is shown
in Figs. \ref{fig:QFT:spectrum1} and \ref{fig:QFT:spectrum2} and is related
to the particle number as in Eq. \eqref{QFT:stat particle number}.
The eigenvalues $\lambda_{h\pm}$ of the exponent of the density operator
are initially infinite (corresponding to zero (anti)particle number),
but oscillate on top of its thermal value of $\lambda_{h\pm}=\beta(\omega_0\mp\mu_1)$
at later time. When more environmental fields are added, the oscillations
are damped and the eigenvalues, and thus the (anti)particle numbers
move closer to a constant. This is similar to what happened in the quantum mechanical
case, see Figs. \ref{fig:50coupledfermions5}--\ref{fig:50coupledfermions4}.

\section{Discussion}
\label{Discussion}

In this work we provide a quantitative description
of the entropy of quantum mechanical and quantum field theoretic fermionic systems,
which here consists of one system oscillator (or field) and $N$ environmental
oscillators (or fields). In our correlator approach to decoherence
the observer is assumed to be sensitive only
to the 2-point correlators of the fermionic
system oscillator (or field). In that case the reduced density operator of the system
is Gaussian and the corresponding Gaussian entropy can be explicitly calculated in terms of
the correlators. We have done this for a one-dimensional fermionic harmonic
oscillator \eqref{HO:entropy:2}, for a fermionic quantum field theory
\eqref{QFT:entropy} and for the general case of $N$ fermionic degrees
of freedom \eqref{NDOF: generalisationEntropy}. We have demonstrated that
the Gaussian density matrix singles out a Fock basis in which it becomes
diagonal. The Fock basis defines statistical particle number, whose
dynamics can be used to define how a system evolves from quantum to classical.
In that sense this Fock basis defines pointer states. An observer which measures
statistical particle number, when it gets entangled with these pointer states,
will experience complete decoherence. The decoherence rate associated with
that observer can be defined to be the classicalization rate.
In Sec.~\ref{sec:Fermionic dynamics} we make a rough estimate
of this rate for the quantum mechanical case
studied here, but we leave a more detailed study of the classicalization rate
in realistic fermionic quantum field theories for future work.

For simplicity in this work we have considered fermionic problems where the system couples
bilinearly to the environment. This problem has the advantage that
it can be solved exactly by numerical methods. We have demonstrated that
the dynamics in general leads to an increase in entropy of the system.
When the system couples strongly to the environment in a thermal
state at temperature $T$, at late times the system's entropy
reaches its thermal value at the same temperature $T$.
Furthermore, in the field theoretic case we have shown that,
when environmental fermionic fields are in a chemical equilibrium
with the common particle chemical potential,
the system field will eventually reach chemical equilibrium with
the environment.

While in this work we focus our attention on the study of exactly soluble
Gaussian systems with bilinear couplings, fermionic systems occuring in
Nature are usually not of that type. An important example is a relativistic
quantum field theory with Yukawa interactions~\eqref{Yukawa}, where
the scalar field is the Standard Model Higgs, a candidate for which
has recently been discovered~\cite{:2012gk,:2012gu},
or the inflaton in inflationary models.
A more sophisticated treatment of
the scalar field in the Yukawa interactions is desirable, and
one can foresee solving nonlinear, perturbative, Kadanoff-Baym
equations for the fermionic and scalar fields, whereby scalar thermal
fluctuations are also taken into account.
Analogous equations have already been tackled within a certain
approximation scheme for the bosonic case in
\cite{Giraud:2009tn,Koksma:2009wa,Koksma:2011dy}.
We intend to address the analogous problem for fermions in future work.

 Another interesting extension of this work would be to study the effects of
CP violation by adding coupling to a pseudo-scalar fermionic current with
time (or space) dependent mass mixing terms (thus emulating phase transitions
in the early Universe). In this case, the CP violation would induce
a difference between the particle and antiparticle numbers,
which in the massless limit becomes the axial vector current.
Studying how this axial current depends on the environmental
temperature in the presence of a non-adiabatically changing mass
would allow for a better understanding of
baryogenesis and leptogenesis sources~\cite{Kainulainen:2001cn,Kainulainen:2002th,Prokopec:2003pj,
Prokopec:2004ic,Konstandin:2003dx,Konstandin:2004gy,Konstandin:2005cd}.

\section{Acknowledgements}

Two authors (TP and JW) were in part supported by the Dutch Foundation for
'Fundamenteel Onderzoek der Materie' (FOM) under the program
"Theoretical particle physics in the era of the LHC", program number FP 104.

\appendix

\section{Bosonic density operator and entropy}
\label{Appendix A: Bosonic density operator and entropy}

In this appendix we calculate the invariant (phase space) area
and entropy for a quantum mechanical system of one bosonic degree
of freedom with position operator $\hat{\phi}$ and momentum operator
$\hat{\pi}$, based on the approach in Ref. \cite{Calzetta:2003dk}.
The {\it Ansatz} for the bosonic density operator is
\begin{equation}
\hat{\rho}_{\rm B}(t)=\frac{1}{Z}
\exp\left[-\frac12(\alpha \hat{\pi}^2+\beta \{\hat{\phi},\hat{\pi}\}+\gamma\hat{\phi}^2)\right]
\label{app1: ansatz bosonic density operator}
\,,
\end{equation}
where $\{.,.\}$ is the anticommutator and $\alpha, \beta, \gamma$ are real time dependent parameters.
By defining bosonic creation and annihilation operators
\begin{align}
\hat{a}_{\rm B}=\sqrt{\frac{\sigma}{2\alpha}}
\left[\left(1+\imath\frac{\beta}{\sigma}\right)\hat{\phi}+\imath\frac{\alpha}{\sigma}\hat{\pi}\right]
\,,\qquad
\hat{a}^{\dagger}_{\rm B}=\sqrt{\frac{\sigma}{2\alpha}}
\left[\left(1-\imath\frac{\beta}{\sigma}\right)\hat{\phi}-\imath\frac{\alpha}{\sigma}\hat{\pi}\right]
\,, \qquad
\sigma \equiv \sqrt{\alpha\gamma-\beta^2}
\,,
\end{align}
the density operator can be written in diagonalised form
\begin{equation}
\hat{\rho}_{\rm B}(t)=\frac{1}{Z'}\exp(-\sigma \hat{N}_{\rm B}),\qquad\qquad Z'\equiv\frac{Z}{{\rm e}^{-\sigma/2}}
\,.
\end{equation}
The bosonic particle number is defined in the usual way
$\hat{N}_{\rm{B}}=\hat{a}^{\dagger}_{\rm{B}}\hat{a}_{\rm{B}}$,
with a corresponding Fock basis $|n_{\rm{B}}\rangle$ defined through
$\hat{N}_{\rm{B}}|n_{\rm{B}}\rangle=n_{\rm{B}}|n_{\rm{B}}\rangle$. Using
this basis to take the trace and demanding that $\rm{Tr}[\hat{\rho}_{\rm{B}}]=1$
we find
\begin{equation}
Z'={\rm{Tr}}[\exp(-\sigma \hat{N}_{\rm{B}})]
=\sum_{n_{\rm{B}}=0}^{\infty}\langle n_{\rm{B}}| \exp(-\sigma \hat{N}_{\rm{B}}) |n_{\rm{B}}\rangle
=\sum_{n_{\rm{B}}=0}^{\infty} {\rm e}^{-\sigma n_{\rm{B}}} = \frac{1}{1-{\rm e}^{-\sigma}}
\,.
\end{equation}
The average particle number is
\begin{equation}
\langle \hat{N}_{\rm{B}}\rangle={\rm{Tr}}[\hat{\rho}_{\rm{B}}\hat{N}_{\rm{B}}]=\frac{1}{{\rm e}^{\sigma}-1}
\equiv \bar{n}_{\rm B}
\,,
\end{equation}
which indeed agrees with the Bose-Einstein distribution
for a thermal state if we identify $\sigma= E/(k_B T)$,
where $k_B$ is the Stefan-Boltzmann constant.
The Gaussian correlators are obtained from $\hat{\rho}_{\rm B}$ as
\begin{align}
\nonumber \langle \hat{\pi}^2 \rangle
  &= -2 \frac{\partial}{\partial\alpha}\ln{Z}
= \Big(\bar{n}_{\rm B}+\frac12\Big)\frac{\gamma}{\sigma}\\
\nonumber
\langle \hat{\phi}^2 \rangle &=\Big(\bar{n}_{\rm B}+\frac12\Big)
                   \frac{\alpha}{\sigma}\\
 \frac12 \langle\{\hat{\phi},\hat{\pi}\} \rangle
          &=\Big(\bar{n}_{\rm B}+\frac12\Big)\frac{-\beta}{\sigma}
\label{app1:bosonic correlators}
\,.
\end{align}
The statistical correlator for bosons is
\begin{equation}
F_{\phi}(t;t')= \frac12\langle\{\hat{\phi}(t),\hat{\phi}(t')\} \rangle
\,,
\end{equation}
which we use to define a Gaussian invariant
\begin{equation}
\Delta_{\phi}(t)=4\left[\langle \hat{\phi}^2 \rangle\langle \hat{\pi}^2\rangle
- \langle \frac12 \{\hat{\phi},\hat{\pi}\} \rangle^2\right]
=4\left.\left[F_{\phi}(t;t')\partial_t\partial_{t'}F_{\phi}(t;t')-(\partial_t F_{\phi}(t;t'))^2\right]\right|_{t=t'}
\label{app1: phase space area}
\,.
\end{equation}
$\Delta_{\phi}/2$ is the phase space area occupied by a Gaussian state in units of $\hbar$
\cite{Koksma:2010dt}.
In a free theory $\Delta_{\phi}=1$ and conserved, whereas it increases for interacting theories.
Using the correlators \eqref{app1:bosonic correlators} we find
\begin{equation}
\Delta_{\phi}(t)=1+2\bar{n}_{\rm B}(t)=\frac{1}{\tanh{\left({\sigma}/{2}\right)}}
\label{app1:bosonic relation phase space area}
\,,
\end{equation}
which presents a relation between the invariant phase space area
and the Gaussian invariant of the density matrix $\sigma$,
and should be compared to the expression for fermions
\eqref{Delta psi}. Finally, the bosonic entropy is
\begin{align}
S_{\phi}=-{\rm Tr}[\ln \hat{\rho}_{\rm B}] = \frac{1+\Delta_\phi}{2}\ln\frac{1+\Delta_\phi}{2}
      -\frac{1-\Delta_\phi}{2}\ln\frac{1-\Delta_\phi}{2}
      =(1+\bar{n}_{\rm B})\ln(1+\bar{n}_{\rm B})-\bar{n}_{\rm B}\ln \bar{n}_{\rm B}
\label{app1:bosonic entropy}
\,.
\end{align}

\section{Fermionic shift and diagonalisation}
\label{Appendix B: Fermionic shift and diagonalisation}

The lagrangian for an interacting fermionic oscillator
is given by (see Eqs. \eqref{HO:entropy:2} and \eqref{current:lagrangian}),
\begin{equation}
 L_\psi = \hat\psi^\dagger (\imath \partial_t-\omega(t))\hat\psi
           - \hat j_\psi^\dagger\hat\psi- \hat\psi^\dagger\hat j_\psi
\,.
\label{APP1:lagrangian}
\end{equation}
This implies the equations of motion,
\begin{equation}
  (\imath \partial_t-\omega(t))\hat\psi = {\hat j}_\psi
\,,\qquad
 (-\imath \partial_t-\omega(t))\hat\psi^\dagger
          = {\hat j}_\psi^\dagger
\,.
\label{APP1:eom}
\end{equation}
One can easily construct the free field solution (in the absence of currents),
\begin{equation}
 \hat\psi_0(t) = {\rm exp}\Big({-\imath \int_0^t\omega d\tau}\Big)\hat\psi_0 (0)
\,,\qquad
 \hat\psi_0^\dagger(t)
         = {\rm exp}\Big({\imath \int_0^t\omega d\tau}\Big)\hat\psi^\dagger_0(0)
\,,
\label{APP1:free fields}
\end{equation}
in terms of which we can express the (free) retarded and advanced
Green functions~(\ref{retarded and advanced Green function}) as,
\begin{equation}
 \imath S_0^{\rm r}(t;t^\prime) = -\imath\theta(t-t^\prime)
                     {\rm e}^{-\imath \int_{t^\prime}^t\omega d\tau}
\,,\quad
 \imath S_0^{\rm a}(t;t^\prime) = \imath\theta(t^\prime-t)
                  {\rm e}^{\imath \int_{t^\prime}^t\omega d\tau}
\label{retarded and advanced Green function:free}
\end{equation}
With a help of $\imath S_0^{\rm r}$ we can solve the general fermionic
operator equations~(\ref{APP1:eom}),
\begin{equation}
 \hat\psi(t) = {\rm e}^{-\imath \int_0^t\omega d\tau}\hat\psi_0(0)
             -\imath\int_0^t dt^\prime
          {\rm e}^{-\imath\int_{t^\prime}^t\omega d\tau} \hat j_\psi(t^\prime)
\,,\qquad
 \hat\psi^\dagger(t)={\rm e}^{\imath \int_0^t\omega d\tau}\hat\psi^\dagger_0(0)
        +\imath\int_0^t dt^\prime
   {\rm e}^{\imath\int_{t^\prime}^t\omega d\tau} \hat j^\dagger_\psi(t^\prime)
\,,
\label{APP1:full solutions}
\end{equation}

 Now, for free fields $\hat\psi_0(t)
 ={\rm e}^{-\imath \int_0^t\omega d\tau}\hat\psi_0(0)$
and the related $\hat\psi_0^\dagger(t)$ we have,
\begin{equation}
 \hat N_0(t) = \hat \psi_0^\dagger(t)\hat\psi_0(t)
             =  \hat\psi_0^\dagger(0) \hat\psi_0(0)
             = \hat N_0(0)
\label{free N}
\end{equation}
which implies that in this (source-free) case the density
matrix~\eqref{HO:rho:solution:2} does not evolve in time,
such that $a=\rm const.$, $\bar n=\rm const.$ and
also the entropy~(\ref{HO:entropy}) $S=\rm const.$

 In the case when there is a nonvanishing current source,
$\hat j_\psi(t)\neq 0$, one can think of the full
solutions~(\ref{APP1:full solutions}) as a suitably shifted $\hat\psi_0(t)$
(analogous to the bosonic Glauber's  coherent states), and therefore
one can write the density operator in terms of
the shifted fields
\begin{equation}
 \hat\psi(t) + \imath\int_0^t dt^\prime
        {\rm e}^{-\imath\int_{t^\prime}^t\omega d\tau} {\hat j}_\psi(t^\prime)
\equiv \hat\psi_0(t)
\,.
\label{shifted field}
\end{equation}
Eq.~(\ref{free N}) then implies that for that density operator
the (von Neumann) entropy is conserved, as it should be.

\section{Exact entropy for two coupled fermions}
\label{Appendix C: Exact entropy for two coupled fermions}

For one environmental oscillator the equations of motion for the statistical correlators
Eqs. \eqref{BCF:eomFqq} become
\begin{align}
\nonumber \imath\partial_t F_{xx}(t;t)&=-\lambda \Delta F(t;t)\\
\nonumber \imath\partial_t\Delta F(t;t)&= (\omega_0-\omega_1)F_{+}(t;t)-2\lambda(F_{xx}(t;t)-F_{qq}(t;t))\\
\nonumber \imath\partial_tF_{+}(t;t)&=(\omega_0-\omega_1)\Delta F(t;t)\\
\imath\partial_tF_{qq}(t;t)&= \lambda\Delta F(t;t)
\,,
\label{App:BCF:eomFqqrewritten}
\end{align}
where $\Delta F(t;t)=F_{xq}(t;t)-F_{qx}(t;t)$ and $F_{+}(t;t)=F_{xq}(t;t)+F_{qx}(t;t)$.
The statistical correlator can be solved from the first line
\begin{equation}
F_{xx}(t;t)=F_{xx}(t_0;t_0)-\frac{\lambda}{\imath}\int_{t_0}^t dt' \Delta F(t';t')
=\frac12 -\frac{\lambda}{\imath}\int_{t_0}^t dt' \Delta F(t';t')
\,,
\label{App:BCF:solutionFxx}
\end{equation}
where we have used the initial conditions \eqref{BCF:initialFxq}.
After acting with $\imath \partial_t$ on the second line of
\eqref{App:BCF:eomFqqrewritten} one finds
\begin{equation}
(\partial_t^2+(\omega_0-\omega_1)^2+4\lambda^2)\Delta F(t;t)=0
\,.
\end{equation}
This equation can be solved with initial conditions for the correlator $\Delta F(t;t)$
itself from Eqs. \eqref{BCF:initialFxq} and for the first time derivative of
the correlator $\Delta F(t;t)$ from the second line of \eqref{App:BCF:eomFqqrewritten}.
This gives
\begin{equation}
\Delta F(t;t)
= \imath \frac{2\lambda}{\bar{\omega}} \bar{n}_{\rm{E}} \sin[\bar{\omega}(t-t_0)]
\,,
\label{App:BCF:solutionDeltaF}
\end{equation}
where
\begin{align}
\nonumber \bar{\omega} &= \sqrt{(\omega_0-\omega_1)^2+4\lambda^2}\\
\bar{n}_{\text{E}}&=\frac{1}{{\rm e}^{\beta \omega_1}+1}
   =\frac12-\frac12\tanh\Big[{\frac12 \beta \omega_1}\Big]
\,.
\end{align}
Inserting the solution \eqref{App:BCF:solutionDeltaF}
in \eqref{App:BCF:solutionFxx}
one obtains the Gaussian invariant from Eq. \eqref{BCF:invareafermionsystem},
\begin{equation}
\Delta_{xx}(t)=1-2\bar{n}_{\text{E}}\left(\frac{2\lambda}{\bar{\omega}}\right)^2
             \sin^2\Big[\frac{\bar{\omega}}{2}(t-t_0)\Big]
\,.
\label{App:BCF:invariantareaexact}
\end{equation}
Thus, the entropy of the system can be analytically calculated
using \eqref{BCF:entropysytem}.

\section{Entropy via the replica trick in coherent state basis}
\label{Appendix D: Entropy via the replica trick in coherent state basis}
Here we calculate the Gaussian von Neumann entropy \eqref{entropy:vN}
using the density operator in the coherent state basis
\eqref{rho in coherent state rep:2}. For convenience we use the exponentiated
form of the elements of the density operator (see Eq. \eqref{HO:rho:solution:3}),
\begin{equation}
\rho(\bar\theta^\prime,\theta;t)=\langle \theta'| \hat{\rho} | \theta \rangle=
\frac{1}{Z}\exp( \bar\theta^\prime M \theta )
\,,
\label{APPD: density operator exponential form}
\end{equation}
where $M=\frac{\bar{n}}{1-\bar{n}}=e^{-a}$ and $Z=\frac{1}{1-\bar{n}}=1+e^{-a}=1+M$.
By making use of the replica trick the entropy can be expressed as
\begin{equation}
S=- {\rm Tr}[\hat \rho\ln(\hat \rho)]=
- \lim_{n\rightarrow 0} \frac{{\rm Tr}[\hat \rho^{n+1}-\hat \rho]}{n}
\,.
\label{APPD: entropy via replica trick}
\end{equation}
The trace is defined in \eqref{coherent basis: trace}.
By inserting $n$ unity operators \eqref{decomposition of unity}
in \eqref{APPD: entropy via replica trick} and using
$\langle \theta^\prime | \theta \rangle = \exp(\theta^{\prime}\theta)$
and Eq. \eqref{APPD: density operator exponential form} one finds
\begin{align}
\nonumber {\rm Tr}[\hat \rho^{n+1}]&=
\int d\theta d\bar{\theta} \exp(\bar\theta \theta)
\prod_{i=1}^{n}\left[\int d\bar\theta^{(i)}d\theta^{(i)}\exp(-\bar\theta^{(i)}\theta^{(i)})\right]
\rho(\bar\theta,\theta^{(1)};t)\times \rho(\bar\theta^{(1)},\theta^{(2)};t)\times ..\times \rho(\bar\theta^{(n)},\theta;t)\\
\nonumber &=Z^{-n-1}\int d\theta d\bar{\theta} \exp(\bar\theta \theta)
\prod_{i=1}^{n}\left[\int d\bar\theta^{(i)}d\theta^{(i)}\exp(-\bar\theta^{(i)}\theta^{(i)})\right]
\exp(\bar\theta M\theta^{(1)})\times\exp(\bar\theta^{(1)}M\theta^{(2)})\times..\times \exp(\bar\theta^{(n)}M\theta)\\
&= Z^{-n-1}(1+M^{n+1})
\,.
\end{align}
The Grassmann integrations have been performed explicitly in the last step.
The resulting entropy \eqref{APPD: entropy via replica trick} becomes
\begin{equation}
S=- \lim_{n\rightarrow 0}\left\{\frac{1+M^{n+1}}{(1+M)^{n+1}}-1\right\}
=-\frac{M}{1+M}\ln{M}+\ln(1+M)
=-(1-\bar{n})\ln(1-\bar{n})-\bar{n}\ln\bar{n}
\,,
\end{equation}
which is indeed the same as the entropy derived
earlier using the Fock basis \eqref{HO:entropy}.

The previous derivation can be generalised for
$N$ fermionic degrees of freedom. In that case
(see Eq. \eqref{NDOF: rhocoherentrepresentation})
\begin{equation}
\rho(\bar{\theta}',\theta;,t)=\langle \theta'| \hat{\rho} | \theta \rangle
= \frac{1}{Z}\exp(\bar{\theta}^{\prime}_i M_{ij} \theta_j)
\,,
\label{APPD: NDOF: Density operator exponential}
\end{equation}
with
\begin{align}
\nonumber M_{ij}&=\left({\rm e}^{-a}\right)_{ij}=\left(\frac{\bar{n}}{1-\bar{n}}\right)_{ij}\\
Z&={\rm{Det}}[\mathbb{I}+{\rm e}^{-a}]={\rm{Det}}[\mathbb{I}+M]
\,.
\end{align}
Again the replica trick \eqref{APPD: entropy via replica trick}
is used to obtain the entropy. The trace is of course taken
over all $N$ fermionic degrees of freedom, which are also included
in the unit operation. After inserting the unity operators
in \eqref{APPD: entropy via replica trick} one finds
\begin{align}
\nonumber {\rm Tr}[\hat \rho^{n+1}]=&
\prod_{a=1}^N\left[\int d\theta_a d\bar{\theta}_a \exp\left(\bar\theta_a \theta_a\right)\right]
\times
\prod_{b=1}^{N}\left[\int d\bar\theta_b^{(1)}d\theta_b^{(1)}\exp\left(-\bar\theta_b^{(1)}\theta_b^{(1)}\right)\right]
\times...\times
\prod_{c=1}^{N}\left[\int d\bar\theta_c^{(n)}d\theta_c^{(n)}\exp\left(-\bar\theta_c^{(n)}\theta_c^{(n)}\right)\right]\\
\nonumber &\times
\rho\left(\bar\theta,\theta^{(1)};t\right)\times \rho\left(\bar\theta^{(1)},\theta^{(2)};t\right)
\times ..\times \rho\left(\bar\theta^{(n)},\theta;t\right)\\
\nonumber =&Z^{-n-1}\prod_{a=1}^N\left[\int d\theta_a d\bar{\theta}_a \exp\left(\bar\theta_a \theta_a\right)\right]
\times
\prod_{b=1}^{N}\left[\int d\bar\theta_b^{(1)}d\theta_b^{(1)}\exp\left(-\bar\theta_b^{(1)}\theta_b^{(1)}\right)\right]
\times...\times
\prod_{c=1}^{N}\left[\int d\bar\theta_c^{(n)}d\theta_c^{(n)}\exp\left(-\bar\theta_c^{(n)}\theta_c^{(n)}\right)\right]\\
& \times
\exp\left(\sum_{i,j}\bar\theta_i M_{ij}\theta_j^{(1)}\right)
\times\exp\left(\sum_{k,l}\bar\theta_k^{(1)}M_{kl}\theta_l^{(2)}\right)
\times..\times
\exp\left(\sum_{r,s}\bar\theta_{r}^{(n)}M_{rs}\theta_s\right)
\,.
\label{APPD: trace rho n+1 N dof}
\end{align}
To avoid confusion the summations have been written
out explicitly. In order to perform the Grassmann integrations
the following identities prove to be useful,
\begin{align}
{\rm Det}[1+M]&=\prod_{a=1}^N\left[\int d\theta_a d\bar{\theta}_a \exp(\bar\theta_a \theta_a)\right]\exp(\sum_{i,j}\bar{\theta}_i M_{ij} \theta_j)\\
\exp(\sum_{i,j}\bar{\theta}_i \left(M^2\right)_{ij} \theta_j)
&= \prod_{b=1}^{N}\left[\int d\bar\theta_b^{(1)}d\theta_b^{(1)}\exp(-\bar\theta_b^{(1)}\theta_b^{(1)})\right]
\exp(\sum_{i,k}\bar\theta_i M_{ik}\theta_k^{(1)})\times\exp(\sum_{l,j}\bar\theta_l^{(1)}M_{lj}\theta_j)
\,.
\end{align}
Applying these to Eq. \eqref{APPD: trace rho n+1 N dof} we find
\begin{equation}
{\rm Tr}[\hat \rho^{n+1}]={\rm Det}[\mathbb{I}+M^{n+1}]
\,,
\end{equation}
The entropy \eqref{APPD: entropy via replica trick} becomes
\begin{equation}
S=- \lim_{n\rightarrow 0}
\left\{\frac{{\rm Det}[\mathbb{I}+M^{n+1}]}{\left({\rm Det}[\mathbb{I}+M]\right)^{n+1}}-1\right\}
={\rm Tr}\left[-\frac{M}{\mathbb{I}+M}\ln{M}+\ln(\mathbb{I}+M)\right]
={\rm Tr}\left[-(1-\bar{n})\ln(1-\bar{n})-\bar{n}\ln\bar{n}\right]
\,,
\end{equation}
which is indeed the entropy derived in Eq. \eqref{NDOF: generalisationEntropy}.
Thus, by using the density operator in the coherent state basis
in combination with the replica trick, no diagonalisation of
the density operator is required in order to find the entropy.

\bibliography{EntropyFermionsBibliography}{}

\begin{thebibliography}{36}
\expandafter\ifx\csname natexlab\endcsname\relax\def\natexlab#1{#1}\fi
\expandafter\ifx\csname bibnamefont\endcsname\relax
  \def\bibnamefont#1{#1}\fi
\expandafter\ifx\csname bibfnamefont\endcsname\relax
  \def\bibfnamefont#1{#1}\fi
\expandafter\ifx\csname citenamefont\endcsname\relax
  \def\citenamefont#1{#1}\fi
\expandafter\ifx\csname url\endcsname\relax
  \def\url#1{\texttt{#1}}\fi
\expandafter\ifx\csname urlprefix\endcsname\relax\def\urlprefix{URL }\fi
\providecommand{\bibinfo}[2]{#2}
\providecommand{\eprint}[2][]{\url{#2}}

\bibitem[{\citenamefont{Giraud and Serreau}(2010)}]{Giraud:2009tn}
\bibinfo{author}{\bibfnamefont{A.}~\bibnamefont{Giraud}} \bibnamefont{and}
  \bibinfo{author}{\bibfnamefont{J.}~\bibnamefont{Serreau}},
  \bibinfo{journal}{Phys.Rev.Lett.} \textbf{\bibinfo{volume}{104}},
  \bibinfo{pages}{230405} (\bibinfo{year}{2010}), \eprint{0910.2570}.

\bibitem[{\citenamefont{Koksma et~al.}(2010{\natexlab{a}})\citenamefont{Koksma,
  Prokopec, and Schmidt}}]{Koksma:2009wa}
\bibinfo{author}{\bibfnamefont{J.~F.} \bibnamefont{Koksma}},
  \bibinfo{author}{\bibfnamefont{T.}~\bibnamefont{Prokopec}}, \bibnamefont{and}
  \bibinfo{author}{\bibfnamefont{M.~G.} \bibnamefont{Schmidt}},
  \bibinfo{journal}{Phys.Rev.} \textbf{\bibinfo{volume}{D81}},
  \bibinfo{pages}{065030} (\bibinfo{year}{2010}{\natexlab{a}}),
  \eprint{0910.5733}.

\bibitem[{\citenamefont{Calzetta and Hu}(1988)}]{Calzetta:1986cq}
\bibinfo{author}{\bibfnamefont{E.}~\bibnamefont{Calzetta}} \bibnamefont{and}
  \bibinfo{author}{\bibfnamefont{B.}~\bibnamefont{Hu}},
  \bibinfo{journal}{Phys.Rev.} \textbf{\bibinfo{volume}{D37}},
  \bibinfo{pages}{2878} (\bibinfo{year}{1988}).

\bibitem[{\citenamefont{Calzetta and Hu}(2003)}]{Calzetta:2003dk}
\bibinfo{author}{\bibfnamefont{E.}~\bibnamefont{Calzetta}} \bibnamefont{and}
  \bibinfo{author}{\bibfnamefont{B.}~\bibnamefont{Hu}},
  \bibinfo{journal}{Phys.Rev.} \textbf{\bibinfo{volume}{D68}},
  \bibinfo{pages}{065027} (\bibinfo{year}{2003}), \eprint{hep-ph/0305326}.

\bibitem[{\citenamefont{Campo and
  Parentani}(2008{\natexlab{a}})}]{Campo:2008ju}
\bibinfo{author}{\bibfnamefont{D.}~\bibnamefont{Campo}} \bibnamefont{and}
  \bibinfo{author}{\bibfnamefont{R.}~\bibnamefont{Parentani}},
  \bibinfo{journal}{Phys.Rev.} \textbf{\bibinfo{volume}{D78}},
  \bibinfo{pages}{065044} (\bibinfo{year}{2008}{\natexlab{a}}),
  \eprint{0805.0548}.

\bibitem[{\citenamefont{Campo and
  Parentani}(2008{\natexlab{b}})}]{Campo:2008ij}
\bibinfo{author}{\bibfnamefont{D.}~\bibnamefont{Campo}} \bibnamefont{and}
  \bibinfo{author}{\bibfnamefont{R.}~\bibnamefont{Parentani}},
  \bibinfo{journal}{Phys.Rev.} \textbf{\bibinfo{volume}{D78}},
  \bibinfo{pages}{065045} (\bibinfo{year}{2008}{\natexlab{b}}),
  \eprint{0805.0424}.

\bibitem[{\citenamefont{Koksma et~al.}(2011{\natexlab{a}})\citenamefont{Koksma,
  Prokopec, and Schmidt}}]{Koksma:2010dt}
\bibinfo{author}{\bibfnamefont{J.~F.} \bibnamefont{Koksma}},
  \bibinfo{author}{\bibfnamefont{T.}~\bibnamefont{Prokopec}}, \bibnamefont{and}
  \bibinfo{author}{\bibfnamefont{M.~G.} \bibnamefont{Schmidt}},
  \bibinfo{journal}{Annals Phys.} \textbf{\bibinfo{volume}{326}},
  \bibinfo{pages}{1548} (\bibinfo{year}{2011}{\natexlab{a}}),
  \eprint{1012.3701}.

\bibitem[{\citenamefont{Koksma et~al.}(2010{\natexlab{b}})\citenamefont{Koksma,
  Prokopec, and Schmidt}}]{Koksma:2010zi}
\bibinfo{author}{\bibfnamefont{J.~F.} \bibnamefont{Koksma}},
  \bibinfo{author}{\bibfnamefont{T.}~\bibnamefont{Prokopec}}, \bibnamefont{and}
  \bibinfo{author}{\bibfnamefont{M.~G.} \bibnamefont{Schmidt}},
  \bibinfo{journal}{Annals Phys.} \textbf{\bibinfo{volume}{325}},
  \bibinfo{pages}{1277} (\bibinfo{year}{2010}{\natexlab{b}}),
  \eprint{1002.0749}.

\bibitem[{\citenamefont{Koksma et~al.}(2012)\citenamefont{Koksma, Prokopec, and
  Schmidt}}]{Koksma:2011fx}
\bibinfo{author}{\bibfnamefont{J.~F.} \bibnamefont{Koksma}},
  \bibinfo{author}{\bibfnamefont{T.}~\bibnamefont{Prokopec}}, \bibnamefont{and}
  \bibinfo{author}{\bibfnamefont{M.~G.} \bibnamefont{Schmidt}},
  \bibinfo{journal}{Phys.Lett.} \textbf{\bibinfo{volume}{B707}},
  \bibinfo{pages}{315} (\bibinfo{year}{2012}), \eprint{1101.5323}.

\bibitem[{\citenamefont{Koksma et~al.}(2011{\natexlab{b}})\citenamefont{Koksma,
  Prokopec, and Schmidt}}]{Koksma:2011dy}
\bibinfo{author}{\bibfnamefont{J.~F.} \bibnamefont{Koksma}},
  \bibinfo{author}{\bibfnamefont{T.}~\bibnamefont{Prokopec}}, \bibnamefont{and}
  \bibinfo{author}{\bibfnamefont{M.~G.} \bibnamefont{Schmidt}},
  \bibinfo{journal}{Phys.Rev.} \textbf{\bibinfo{volume}{D83}},
  \bibinfo{pages}{085011} (\bibinfo{year}{2011}{\natexlab{b}}),
  \eprint{1102.4713}.

\bibitem[{\citenamefont{Floreanini and Jackiw}(1988)}]{Floreanini:1987gr}
\bibinfo{author}{\bibfnamefont{R.}~\bibnamefont{Floreanini}} \bibnamefont{and}
  \bibinfo{author}{\bibfnamefont{R.}~\bibnamefont{Jackiw}},
  \bibinfo{journal}{Phys.Rev.} \textbf{\bibinfo{volume}{D37}},
  \bibinfo{pages}{2206} (\bibinfo{year}{1988}).

\bibitem[{\citenamefont{Kiefer and Wipf}(1994)}]{Kiefer:1993fw}
\bibinfo{author}{\bibfnamefont{C.}~\bibnamefont{Kiefer}} \bibnamefont{and}
  \bibinfo{author}{\bibfnamefont{A.}~\bibnamefont{Wipf}},
  \bibinfo{journal}{Annals Phys.} \textbf{\bibinfo{volume}{236}},
  \bibinfo{pages}{241} (\bibinfo{year}{1994}), \eprint{hep-th/9306161}.

\bibitem[{\citenamefont{Benatti and Floreanini}(2000)}]{Benatti:2000wu}
\bibinfo{author}{\bibfnamefont{F.}~\bibnamefont{Benatti}} \bibnamefont{and}
  \bibinfo{author}{\bibfnamefont{R.}~\bibnamefont{Floreanini}},
  \bibinfo{journal}{J.Phys.A} \textbf{\bibinfo{volume}{A33}},
  \bibinfo{pages}{8139} (\bibinfo{year}{2000}), \eprint{hep-th/0010013}.

\bibitem[{\citenamefont{Zeh}(1970)}]{Zeh:1970}
\bibinfo{author}{\bibfnamefont{H.}~\bibnamefont{Zeh}},
  \bibinfo{journal}{Foundations of Physics} \textbf{\bibinfo{volume}{1}},
  \bibinfo{pages}{69} (\bibinfo{year}{1970}), ISSN \bibinfo{issn}{0015-9018}.

\bibitem[{\citenamefont{Zurek}(1981)}]{Zurek:1981xq}
\bibinfo{author}{\bibfnamefont{W.}~\bibnamefont{Zurek}},
  \bibinfo{journal}{Phys.Rev.} \textbf{\bibinfo{volume}{D24}},
  \bibinfo{pages}{1516} (\bibinfo{year}{1981}).

\bibitem[{\citenamefont{Joos and Zeh}(1985)}]{Joos:1984uk}
\bibinfo{author}{\bibfnamefont{E.}~\bibnamefont{Joos}} \bibnamefont{and}
  \bibinfo{author}{\bibfnamefont{H.}~\bibnamefont{Zeh}},
  \bibinfo{journal}{Z.Phys.} \textbf{\bibinfo{volume}{B59}},
  \bibinfo{pages}{223} (\bibinfo{year}{1985}).

\bibitem[{\citenamefont{Joos et~al.}(1996, second edition
  2003)\citenamefont{Joos, Zeh, Kiefer, Giulini, Kupsch, and
  Stamatescu}}]{joos2003decoherence}
\bibinfo{author}{\bibfnamefont{E.}~\bibnamefont{Joos}},
  \bibinfo{author}{\bibfnamefont{H.}~\bibnamefont{Zeh}},
  \bibinfo{author}{\bibfnamefont{C.}~\bibnamefont{Kiefer}},
  \bibinfo{author}{\bibfnamefont{D.}~\bibnamefont{Giulini}},
  \bibinfo{author}{\bibfnamefont{J.}~\bibnamefont{Kupsch}}, \bibnamefont{and}
  \bibinfo{author}{\bibfnamefont{I.}~\bibnamefont{Stamatescu}},
  \emph{\bibinfo{title}{Decoherence and the Appearance of a Classical World in
  Quantum Theory}} (\bibinfo{publisher}{Springer}, \bibinfo{year}{1996, second
  edition 2003}), ISBN \bibinfo{isbn}{9783540003908},
  \urlprefix\url{http://books.google.nl/books?id=6eTHcxeNxdUC}.

\bibitem[{\citenamefont{Zurek}(2003)}]{Zurek:2003zz}
\bibinfo{author}{\bibfnamefont{W.~H.} \bibnamefont{Zurek}},
  \bibinfo{journal}{Rev.Mod.Phys.} \textbf{\bibinfo{volume}{75}},
  \bibinfo{pages}{715} (\bibinfo{year}{2003}).

\bibitem[{\citenamefont{Eisert et~al.}(2010)\citenamefont{Eisert, Cramer, and
  Plenio}}]{Eisert:2008ur}
\bibinfo{author}{\bibfnamefont{J.}~\bibnamefont{Eisert}},
  \bibinfo{author}{\bibfnamefont{M.}~\bibnamefont{Cramer}}, \bibnamefont{and}
  \bibinfo{author}{\bibfnamefont{M.}~\bibnamefont{Plenio}},
  \bibinfo{journal}{Rev.Mod.Phys.} \textbf{\bibinfo{volume}{82}},
  \bibinfo{pages}{277} (\bibinfo{year}{2010}), \eprint{0808.3773}.

\bibitem[{\citenamefont{Solodukhin}(2011)}]{Solodukhin:2011gn}
\bibinfo{author}{\bibfnamefont{S.~N.} \bibnamefont{Solodukhin}},
  \bibinfo{journal}{Living Rev.Rel.} \textbf{\bibinfo{volume}{14}},
  \bibinfo{pages}{8} (\bibinfo{year}{2011}), \eprint{1104.3712}.

\bibitem[{\citenamefont{Cabibbo}(1963)}]{Cabibbo:1963yz}
\bibinfo{author}{\bibfnamefont{N.}~\bibnamefont{Cabibbo}},
  \bibinfo{journal}{Phys.Rev.Lett.} \textbf{\bibinfo{volume}{10}},
  \bibinfo{pages}{531} (\bibinfo{year}{1963}).

\bibitem[{\citenamefont{Kobayashi and Maskawa}(1973)}]{Kobayashi:1973fv}
\bibinfo{author}{\bibfnamefont{M.}~\bibnamefont{Kobayashi}} \bibnamefont{and}
  \bibinfo{author}{\bibfnamefont{T.}~\bibnamefont{Maskawa}},
  \bibinfo{journal}{Prog.Theor.Phys.} \textbf{\bibinfo{volume}{49}},
  \bibinfo{pages}{652} (\bibinfo{year}{1973}).

\bibitem[{\citenamefont{Pontecorvo}(1968)}]{Pontecorvo:1967fh}
\bibinfo{author}{\bibfnamefont{B.}~\bibnamefont{Pontecorvo}},
  \bibinfo{journal}{Sov.Phys.JETP} \textbf{\bibinfo{volume}{26}},
  \bibinfo{pages}{984} (\bibinfo{year}{1968}).

\bibitem[{\citenamefont{Maki et~al.}(1962)\citenamefont{Maki, Nakagawa, and
  Sakata}}]{Maki:1962mu}
\bibinfo{author}{\bibfnamefont{Z.}~\bibnamefont{Maki}},
  \bibinfo{author}{\bibfnamefont{M.}~\bibnamefont{Nakagawa}}, \bibnamefont{and}
  \bibinfo{author}{\bibfnamefont{S.}~\bibnamefont{Sakata}},
  \bibinfo{journal}{Prog.Theor.Phys.} \textbf{\bibinfo{volume}{28}},
  \bibinfo{pages}{870} (\bibinfo{year}{1962}).

\bibitem[{\citenamefont{Kubo}(1990)}]{kubo1990statistical}
\bibinfo{author}{\bibfnamefont{R.}~\bibnamefont{Kubo}},
  \emph{\bibinfo{title}{Statistical mechanics}}, North-Holland Personal Library
  (\bibinfo{publisher}{North-Holland}, \bibinfo{year}{1990}), ISBN
  \bibinfo{isbn}{9780444871039},
  \urlprefix\url{http://books.google.nl/books?id=b-g9AQAAIAAJ}.

\bibitem[{\citenamefont{Blaizot and Ripka}(1986)}]{blaizot1986quantum}
\bibinfo{author}{\bibfnamefont{J.}~\bibnamefont{Blaizot}} \bibnamefont{and}
  \bibinfo{author}{\bibfnamefont{G.}~\bibnamefont{Ripka}},
  \emph{\bibinfo{title}{Quantum Theory of Finite Systems}}
  (\bibinfo{publisher}{Mit Press}, \bibinfo{year}{1986}), ISBN
  \bibinfo{isbn}{9780262022149},
  \urlprefix\url{http://books.google.nl/books?id=s\_xlQgAACAAJ}.

\bibitem[{\citenamefont{Glauber}(1963)}]{Glauber:1963tx}
\bibinfo{author}{\bibfnamefont{R.~J.} \bibnamefont{Glauber}},
  \bibinfo{journal}{Phys.Rev.} \textbf{\bibinfo{volume}{131}},
  \bibinfo{pages}{2766} (\bibinfo{year}{1963}).

\bibitem[{\citenamefont{Aad et~al.}(2012)}]{:2012gk}
\bibinfo{author}{\bibfnamefont{G.}~\bibnamefont{Aad}} \bibnamefont{et~al.}
  (\bibinfo{collaboration}{ATLAS Collaboration}) (\bibinfo{year}{2012}),
  \eprint{1207.7214}.

\bibitem[{\citenamefont{Chatrchyan et~al.}(2012)}]{:2012gu}
\bibinfo{author}{\bibfnamefont{S.}~\bibnamefont{Chatrchyan}}
  \bibnamefont{et~al.} (\bibinfo{collaboration}{CMS Collaboration}),
  \bibinfo{journal}{Phys.Lett.B}  (\bibinfo{year}{2012}), \eprint{1207.7235}.

\bibitem[{\citenamefont{Kainulainen et~al.}(2001)\citenamefont{Kainulainen,
  Prokopec, Schmidt, and Weinstock}}]{Kainulainen:2001cn}
\bibinfo{author}{\bibfnamefont{K.}~\bibnamefont{Kainulainen}},
  \bibinfo{author}{\bibfnamefont{T.}~\bibnamefont{Prokopec}},
  \bibinfo{author}{\bibfnamefont{M.~G.} \bibnamefont{Schmidt}},
  \bibnamefont{and}
  \bibinfo{author}{\bibfnamefont{S.}~\bibnamefont{Weinstock}},
  \bibinfo{journal}{JHEP} \textbf{\bibinfo{volume}{0106}}, \bibinfo{pages}{031}
  (\bibinfo{year}{2001}), \eprint{hep-ph/0105295}.

\bibitem[{\citenamefont{Kainulainen et~al.}(2002)\citenamefont{Kainulainen,
  Prokopec, Schmidt, and Weinstock}}]{Kainulainen:2002th}
\bibinfo{author}{\bibfnamefont{K.}~\bibnamefont{Kainulainen}},
  \bibinfo{author}{\bibfnamefont{T.}~\bibnamefont{Prokopec}},
  \bibinfo{author}{\bibfnamefont{M.~G.} \bibnamefont{Schmidt}},
  \bibnamefont{and}
  \bibinfo{author}{\bibfnamefont{S.}~\bibnamefont{Weinstock}},
  \bibinfo{journal}{Phys.Rev.} \textbf{\bibinfo{volume}{D66}},
  \bibinfo{pages}{043502} (\bibinfo{year}{2002}), \eprint{hep-ph/0202177}.

\bibitem[{\citenamefont{Prokopec
  et~al.}(2004{\natexlab{a}})\citenamefont{Prokopec, Schmidt, and
  Weinstock}}]{Prokopec:2003pj}
\bibinfo{author}{\bibfnamefont{T.}~\bibnamefont{Prokopec}},
  \bibinfo{author}{\bibfnamefont{M.~G.} \bibnamefont{Schmidt}},
  \bibnamefont{and}
  \bibinfo{author}{\bibfnamefont{S.}~\bibnamefont{Weinstock}},
  \bibinfo{journal}{Annals Phys.} \textbf{\bibinfo{volume}{314}},
  \bibinfo{pages}{208} (\bibinfo{year}{2004}{\natexlab{a}}),
  \eprint{hep-ph/0312110}.

\bibitem[{\citenamefont{Prokopec
  et~al.}(2004{\natexlab{b}})\citenamefont{Prokopec, Schmidt, and
  Weinstock}}]{Prokopec:2004ic}
\bibinfo{author}{\bibfnamefont{T.}~\bibnamefont{Prokopec}},
  \bibinfo{author}{\bibfnamefont{M.~G.} \bibnamefont{Schmidt}},
  \bibnamefont{and}
  \bibinfo{author}{\bibfnamefont{S.}~\bibnamefont{Weinstock}},
  \bibinfo{journal}{Annals Phys.} \textbf{\bibinfo{volume}{314}},
  \bibinfo{pages}{267} (\bibinfo{year}{2004}{\natexlab{b}}).

\bibitem[{\citenamefont{Konstandin et~al.}(2004)\citenamefont{Konstandin,
  Prokopec, and Schmidt}}]{Konstandin:2003dx}
\bibinfo{author}{\bibfnamefont{T.}~\bibnamefont{Konstandin}},
  \bibinfo{author}{\bibfnamefont{T.}~\bibnamefont{Prokopec}}, \bibnamefont{and}
  \bibinfo{author}{\bibfnamefont{M.~G.} \bibnamefont{Schmidt}},
  \bibinfo{journal}{Nucl.Phys.} \textbf{\bibinfo{volume}{B679}},
  \bibinfo{pages}{246} (\bibinfo{year}{2004}), \eprint{hep-ph/0309291}.

\bibitem[{\citenamefont{Konstandin et~al.}(2005)\citenamefont{Konstandin,
  Prokopec, and Schmidt}}]{Konstandin:2004gy}
\bibinfo{author}{\bibfnamefont{T.}~\bibnamefont{Konstandin}},
  \bibinfo{author}{\bibfnamefont{T.}~\bibnamefont{Prokopec}}, \bibnamefont{and}
  \bibinfo{author}{\bibfnamefont{M.~G.} \bibnamefont{Schmidt}},
  \bibinfo{journal}{Nucl.Phys.} \textbf{\bibinfo{volume}{B716}},
  \bibinfo{pages}{373} (\bibinfo{year}{2005}), \eprint{hep-ph/0410135}.

\bibitem[{\citenamefont{Konstandin et~al.}(2006)\citenamefont{Konstandin,
  Prokopec, Schmidt, and Seco}}]{Konstandin:2005cd}
\bibinfo{author}{\bibfnamefont{T.}~\bibnamefont{Konstandin}},
  \bibinfo{author}{\bibfnamefont{T.}~\bibnamefont{Prokopec}},
  \bibinfo{author}{\bibfnamefont{M.~G.} \bibnamefont{Schmidt}},
  \bibnamefont{and} \bibinfo{author}{\bibfnamefont{M.}~\bibnamefont{Seco}},
  \bibinfo{journal}{Nucl.Phys.} \textbf{\bibinfo{volume}{B738}},
  \bibinfo{pages}{1} (\bibinfo{year}{2006}), \eprint{hep-ph/0505103}.

\end{thebibliography}

\end{document}